\documentclass[twocolumn,tighten]{aastex63}

\usepackage{graphicx}
\usepackage{amsmath}
\usepackage{comment}
\usepackage{array,multirow,color,tablefootnote}
\usepackage{xcolor}
\usepackage{subfigure}

\newcommand{\sig}[1]{\ensuremath{#1\sigma}}

\newcommand{\tess}{TESS}

\shorttitle{KELT-9{\rm b} \tess\ Phase Curve}
\shortauthors{Wong et~al.}

\begin{document}
\title{Exploring the Atmospheric Dynamics of the Extreme Ultrahot Jupiter KELT-9b Using TESS photometry} 


\author[0000-0001-9665-8429]{Ian~Wong}
\altaffiliation{51 Pegasi b Fellow}
\affil{Department of Earth, Atmospheric, and Planetary Sciences, Massachusetts Institute of Technology,
Cambridge, MA 02139, USA}

\author[0000-0002-1836-3120]{Avi~Shporer}
\affiliation{Department of Physics and Kavli Institute for Astrophysics and Space Research, Massachusetts Institute of Technology, Cambridge, MA 02139, USA}

\author[0000-0003-4269-3311]{Daniel~Kitzmann}
\affiliation{University of Bern, Center for Space and Habitability, Bern, Switzerland}

\author[0000-0003-2528-3409]{Brett~M.~Morris}
\affiliation{University of Bern, Center for Space and Habitability, Bern, Switzerland}

\author[0000-0003-1907-5910]{Kevin~Heng}
\affiliation{University of Bern, Center for Space and Habitability, Bern, Switzerland}

\author[0000-0001-8981-6759]{H.~Jens~Hoeijmakers}
\affiliation{University of Bern, Center for Space and Habitability, Bern, Switzerland}
\affiliation{Observatoire astronomique de l'Universit{\' e} de Gen{\` e}ve, Geneva, Switzerland}

\author[0000-0002-9355-5165]{Brice-Olivier~Demory}
\affiliation{University of Bern, Center for Space and Habitability, Bern, Switzerland}

\author[0000-0003-2086-7712]{John~P.~Ahlers}
\altaffiliation{NASA Postdoctoral Program Fellow}
\affiliation{NASA Goddard Space Flight Center, 8800 Greenbelt Road, Greenbelt, MD 20771, USA}

\author[0000-0003-4241-7413]{Megan~Mansfield}
\affiliation{Department of Geophysical Sciences, University of Chicago, Chicago, IL 60637, USA}

\author[0000-0003-4733-6532]{Jacob~L.~Bean}
\affiliation{Department of Astronomy \& Astrophysics, University of Chicago, Chicago, IL 60637, USA}

\author[0000-0002-6939-9211]{Tansu~Daylan}
\altaffiliation{Kavli Fellow}
\affiliation{Department of Physics and Kavli Institute for Astrophysics and Space Research, Massachusetts Institute of Technology, Cambridge, MA 02139, USA}

\author{Tara~Fetherolf}
\affiliation{Department of Physics and Astronomy, University of California, Riverside, CA 92521, USA}

\author[0000-0001-8812-0565]{Joseph~E.~Rodriguez}
\affiliation{Center for Astrophysics$\vert$Harvard \& Smithsonian, 60 Garden Street, Cambridge, MA 02138, USA}

\author[0000-0001-5578-1498]{Bj{\" o}rn~Benneke}
\affiliation{Department of Physics and Institute for Research on Exoplanets, Universit{\' e} de Montr{\' e}al, Montr{\' e}al, QC, Canada}

\author[0000-0003-2058-6662]{George~R.~Ricker}
\affiliation{Department of Physics and Kavli Institute for Astrophysics and Space Research, Massachusetts Institute of Technology, Cambridge, MA 02139, USA}

\author[0000-0001-9911-7388]{David~W.~Latham}
\affiliation{Center for Astrophysics$\vert$Harvard \& Smithsonian, 60 Garden Street, Cambridge, MA 02138, USA}

\author[0000-0001-6763-6562]{Roland~Vanderspek}
\affiliation{Department of Physics and Kavli Institute for Astrophysics and Space Research, Massachusetts Institute of Technology, Cambridge, MA 02139, USA}

\author[0000-0002-6892-6948]{Sara~Seager}
\affiliation{Department of Earth, Atmospheric, and Planetary Sciences, Massachusetts Institute of Technology, Cambridge, MA 02139, USA}
\affiliation{Department of Physics and Kavli Institute for Astrophysics and Space Research, Massachusetts Institute of Technology, Cambridge, MA 02139, USA}
\affiliation{Department of Aeronautics and Astronautics, MIT, 77 Massachusetts Avenue, Cambridge, MA 02139, USA}

\author[0000-0002-4265-047X]{Joshua~N.~Winn}
\affiliation{Department of Astrophysical Sciences, Princeton University, Princeton, NJ 08544, USA}

\author[0000-0002-4715-9460]{Jon~M.~Jenkins}
\affiliation{NASA Ames Research Center, Moffett Field, CA 94035, USA}

\author[0000-0002-7754-9486]{Christopher~J.~Burke}
\affiliation{Department of Physics and Kavli Institute for Astrophysics and Space Research, Massachusetts Institute of Technology, Cambridge, MA 02139, USA}

\author[0000-0002-8035-4778]{Jessie~L.~Christiansen}
\affiliation{Caltech/IPAC-NASA Exoplanet Science Institute, Pasadena, CA 91106, USA}

\author[0000-0002-2482-0180]{Zahra~Essack}
\affiliation{Department of Earth, Atmospheric, and Planetary Sciences, Massachusetts Institute of Technology, Cambridge, MA 02139, USA}

\author{Mark~E.~Rose}
\affiliation{NASA Ames Research Center, Moffett Field, CA 94035, USA}

\author[0000-0002-6148-7903]{Jeffrey~C.~Smith}
\affiliation{NASA Ames Research Center, Moffett Field, CA 94035, USA}
\affiliation{SETI Institute, Mountain View, CA 94043, USA}

\author[0000-0002-1949-4720]{Peter~Tenenbaum}
\affiliation{NASA Ames Research Center, Moffett Field, CA 94035, USA}
\affiliation{SETI Institute, Mountain View, CA 94043, USA}

\author[0000-0003-4755-584X]{Daniel~Yahalomi}
\affiliation{Center for Astrophysics$\vert$Harvard \& Smithsonian, 60 Garden Street, Cambridge, MA 02138, USA}

\begin{abstract}
We carry out a phase-curve analysis of the KELT-9 system using photometric observations from NASA's Transiting Exoplanet Survey Satellite (TESS). The measured secondary eclipse depth and peak-to-peak atmospheric brightness modulation are $650^{+14}_{-15}$~ppm and $566\pm16$~ppm, respectively. The planet's brightness variation reaches maximum $31\pm5$~minutes before the midpoint of the secondary eclipse, indicating a $5\overset{\circ}{.}2\pm0\overset{\circ}{.}9$ eastward shift in the dayside hot spot from the substellar point. We also detect stellar pulsations on KELT-9 with a period of $7.58695\pm0.00091$~hours. The dayside emission of KELT-9b in the \tess\ bandpass is consistent with a blackbody brightness temperature of $4600\pm100$~K. The corresponding nightside brightness temperature is $3040\pm100$~K, comparable to the dayside temperatures of the hottest known exoplanets. In addition, we detect a significant phase-curve signal at the first harmonic of the orbital frequency and a marginal signal at the second harmonic. While the amplitude of the first harmonic component is consistent with the predicted ellipsoidal distortion modulation assuming equilibrium tides, the phase of this photometric variation is shifted relative to the expectation. Placing KELT-9b in the context of other exoplanets with phase-curve observations, we find that the elevated nightside temperature and relatively low day--night temperature contrast agree with the predictions of atmospheric models that include H$_{2}$ dissociation and recombination. The nightside temperature of KELT-9b implies an atmospheric composition containing about 50\% molecular and 50\% atomic hydrogen at 0.1~bar, a nightside emission spectrum that deviates significantly from a blackbody, and a 0.5--2.0~$\mu$m transmission spectrum that is featureless at low resolution.
\end{abstract}



\section{Introduction}
\label{sec:intro}

We present the Transiting Exoplanet Survey Satellite (TESS) phase curve of KELT-9b. This 2.9~$M_{J}$ planet lies on a near-polar 1.48-day orbit around a massive, rapidly rotating A0/B9 star ($M=2.3~M_{\Sun}$, $R = 2.4~R_{\Sun}$, $T_{\mathrm{eff}}=$ 10,170~K) and has a dayside equilibrium temperature of $\sim$4600~K, similar to that of a mid-K star \citep{gaudi2017} and more than 1500~K higher than the dayside temperature of the next hottest known planet. KELT-9b lies at the extreme high-temperature end of the so-called ultrahot Jupiters (UHJs), which are emerging as a distinct class of short-period gas giants. These planets have dayside temperatures exceeding $T_{\mathrm{day}}=2200$~K \citep[e.g.,][]{bell2018,parmentier2018} and are characterized by unique physical processes shaping their atmospheric composition and dynamics that are not found on cooler planets.

The UHJs lie in an extreme irradiation regime where thermal dissociation of molecules produces a dayside atmosphere dominated by atomic gases \citep[e.g.,][]{arcangeli2018,hoeijmakers2018,hoeijmakers2019}. Here the combined effects of the continuum opacity of H$^{-}$ \citep{bell2018,kitzmann2018,lothringer2018,parmentier2018}, and the muted H$_{2}$O absorption features formed by strong vertical gradients in the thermal dissociation fraction \citep{kreidberg2018,parmentier2018} are predicted to yield a featureless blackbody-like near-infrared emission spectra at low resolution, as has been measured for several UHJs, such as HAT-P-7b \citep{mansfield2018}, WASP-18b \citep{arcangeli2018}, and WASP-103b \citep{kreidberg2018}. In addition, as molecules responsible for radiative cooling (e.g., H$_{2}$O) are destroyed by the high dayside temperature \citep{kitzmann2018,parmentier2018}, refractory elements such as Fe and Mg continue to absorb incident starlight at UV and optical wavelengths and heat the atmosphere at low pressures, inducing a significant dayside temperature inversion \citep{kitzmann2018,lothringer2018}. Previous observations of KELT-9b have revealed many characteristics that are hallmarks of UHJs, including an extended, escaping atmosphere of dissociated hydrogen \citep{yan2018} and spectral features from ionized Fe, Ti, Mg, Ca, Na, Cr, Sc, and Y in optical transmission spectroscopy \citep{hoeijmakers2018,yan2018,hoeijmakers2019,cauley2019,turner2020}.

Phase-curve measurements provide a first-order picture of a planet's longitudinal brightness distribution \citep[e.g.,][]{cowan2008,shporer2017}. At wavelengths where the thermal emission of the planet dominates, the amplitude and phase shift of the phase-curve variation allow us to measure the day--night brightness temperature contrast and the dayside hot-spot offset relative to the substellar point \citep[e.g.,][]{showman2013,hengshowman2015,parmentier2017}. Both of these properties reflect the efficiency of heat transport in the atmosphere, and as the number of measured exoplanet phase curves has grown dramatically over the past decade alongside improvements in atmospheric modeling, several salient trends have emerged. General circulation models of exoplanet atmospheres predict an increase in the day--night temperature contrast and a corresponding decrease in the dayside hot-spot offset as the level of incident stellar irradiation increases \citep[e.g.][]{perna2012,perezbecker2013,komacek2016}. 

While this trend has been shown to hold for moderately irradiated ($T_{\mathrm{day}}<2500$~K) hot Jupiters \citep[][]{schwartz2017,zhang2018,keating2019}, several UHJs deviate from the expected behavior. Recent phase-curve measurements of WASP-33b \citep{zhang2018} and WASP-103b \citep{kreidberg2018} indicate relatively small day--night temperature contrasts when compared to their cooler counterparts. It has been suggested that the transport of thermally dissociated atomic hydrogen to the nightside and subsequent recombination into H$_{2}$ release a significant amount of heat, which serves to moderate the temperature contrast between the two hemispheres \citep{bell2018,komacek2018,tan2019}. These models predict a rollover in the relationship between day--night temperature contrast and dayside equilibrium temperature at around 2500~K, above which the effect of hydrogen recombination amplifies heat recirculation. 

From an empirical study of Spitzer phase curves, \citet{zhang2018} suggested an analogous rollover in the trend of measured dayside hot-spot offsets, with the hottest UHJs displaying somewhat elevated offset amplitudes relative to cooler planets. However, there is significant scatter in the relationship between dayside hot-spot offsets and irradiation, with some UHJs, such as WASP-103b, showing no evidence for a dayside hot-spot offset \citep{kreidberg2018}. Wavelength coverage introduces an extra dimension to the study of atmospheric dynamics through phase curves. Previous systematic analyses of visible-light phase curves from the Kepler mission have revealed a tentative trend in the direction of the offset in the dayside brightness maximum: cooler planets show offsets to the west of the substellar point, suggesting reflective clouds near the morning terminator, while hotter planets show eastward offsets, consistent with the aforementioned advection of the hot spot due to heat recirculation \citep[][]{esteves2015}. These results suggest a complex interplay between the longitudinal temperature distribution, inhomogeneous clouds, and the systematic trend between the increasing relative contribution of the planet's thermal emission at optical wavelengths and increasing dayside temperature, necessitating further observational and theoretical study.

TESS has proven to be a powerful tool for time-domain exoplanet science. The high temperatures of UHJs make them particularly amenable to detailed phase-curve analysis, because their thermal emission can be easily detected even at optical wavelengths. \tess\ phase curves have been published for WASP-18b \citep{shporer2019}, WASP-19b \citep{wong2019phase}, and WASP-121b \citep{bourrier2019,daylan2019}.

By measuring the secondary eclipse and phase-curve variation of the KELT-9 system in the \tess\ bandpass, we characterize the longitudinal temperature distribution and atmospheric heat transport on this exceptional planet and place our results in the context of other UHJs. We combine our results with those from a recent analysis of the full-orbit 4.5~$\mu$m Spitzer phase curve \citep{mansfield2019}. In addition, we search for the photometric modulation due to ellipsoidal tidal distortion of the host star and compare the fitted amplitude and phase of this variability signal with the predictions from stellar astrophysics.

The structure of the paper is as follows. In Section~\ref{sec:obsana}, we briefly describe the \tess\ observations and data analysis methodologies. The results of the phase-curve analysis are presented in Section~\ref{sec:res}, and in Section~\ref{sec:dis}, we explore the implications of our findings in relation to the dayside and nightside temperature contrast, atmospheric dynamics, the predicted stellar tidal distortion, and overall trends in exoplanet phase curves. A short summary is provided in Section~\ref{sec:con}.

\section{Observations and Data Analysis}
\label{sec:obsana}

\begin{figure*}
\includegraphics[width=\linewidth]{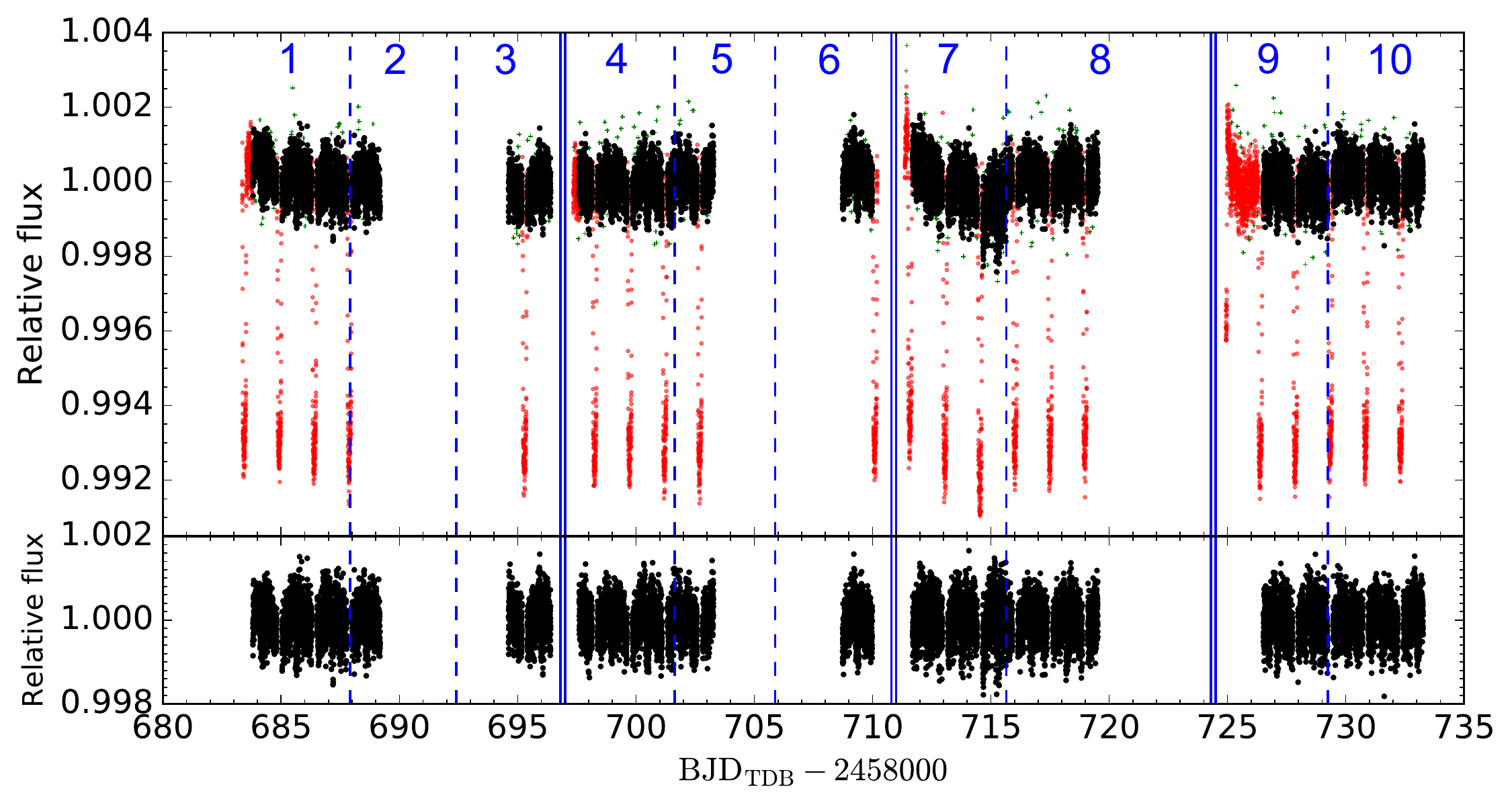}
\singlespace
\caption{Top panel: the normalized PDC light curve of KELT-9 from \tess\ Sectors 14 and 15. The large gaps in data correspond to episodes of severe stray-light contamination on the detector. The vertical blue dashed lines indicate the momentum dumps that occurred during each of the two spacecraft orbits, which separate the light curve into 10 segments (labeled 1--10). The pairs of solid blue lines indicate the boundaries between spacecraft orbits. The red points demarcate the transits, as well as the data segments at the start of each spacecraft orbit during which short-term flux ramps occurred. The green plus signs denote the outliers removed by our moving median filter. All of the red and green data points were trimmed prior to fitting. Bottom panel: the corresponding systematics-removed light curve from our joint fit. The secondary eclipses and atmospheric brightness modulation are clearly discernible.}
\label{fig:lc} 
\end{figure*}

\subsection{\tess\ Light Curve}\label{subsec:data}

The KELT-9 system was observed by camera 1 of the \tess\ spacecraft during Sector 14 (UT 2019 July 18 to August 14) and Sector 15 (UT 2019 August 15 to September 11). KELT-9 is listed in the \tess\ input catalog (TIC; \citealt{stassun2018}) as TIC 16740101 and is included in the list of preselected target stars that are observed with a 2 minute cadence using an 11$\times$11 pixel subarray centered on the target. The photometric data were processed through the Science Processing Operations Center (SPOC) pipeline \citep{jenkins2016}, hosted at the NASA Ames Research Center, which is largely based on the predecessor Kepler mission pipeline \citep{jenkins2010, jenkins2017}. 


Even after instrumental systematics corrections by the SPOC pipeline, there are residual long-term trends in the photometry. We corrected for these by simultaneously fitting generalized polynomials in time to each data segment alongside the astrophysical phase-curve model. The order of each segment's polynomial detrending function was optimized by minimizing the Bayesian information criterion (BIC); across the 10 segments, the optimal orders ranged from zero to 2.

After downloading the data from the Mikulski Archive for Space Telescopes (MAST), we proceeded to analyze the light curves following methodologies that are largely identical to the ones described in previous \tess\ phase-curve papers \citep{shporer2019,wong2019phase}. In the analysis presented in this paper, we used the presearch data conditioning (PDC; \citealt{smith2012, stumpe2014}) light curves from the SPOC pipeline, which have been corrected for instrumental systematics and contamination from nearby stars. We also experimented with fitting the uncorrected simple aperture photometry (SAP) light curves. The SAP data show more significant long-term systematic trends. Correcting these trends using linear combinations of the cotrending basis vectors (CBVs) provided by MAST removed a majority of the systematics and yielded phase-curve fit parameters that are consistent with those derived using the PDC light curves. However, some long-term temporal trends persist, and the corrected SAP light curves have larger scatter than the corrected PDC light curves, resulting in less precise fitted astrophysical parameters. Likewise, we constructed light curves from the target pixel files and defined various extraction apertures but found that the quality of the detrending (using either CBVs or polynomials in time) was generally significantly poorer than in the case of the SPOC-generated SAP light curve.

Data obtained during Sectors 14 and 15 were occasionally affected by severe stray-light contamination on the detector. The SPOC pipeline flagged those exposures and set the fluxes as NaN. In addition, data points preceding and following the episodes of stray-light contamination exhibited higher-than-normal background fluxes and were also flagged in the PDC data. Out of 19,337 2 minute frames obtained by \tess\ in Sector 14, 8603 ($44\%$) were flagged, the vast majority due to stray light; in Sector 15, $37\%$ of the data points (6973 out of 18,757) were flagged. We removed all flagged data points from the light curve and then applied a 16-point moving median filter to the photometric time series to trim \sig{3} outliers. The outlier trimming removed an additional $1.6\%$ (172 out of 10,734) and $1.7\%$ (195 out of 11,784) of the points from the Sector 14 and Sector 15 time series, respectively.

The transits of KELT-9b show significant deviations from the typical transit light-curve shape. These asymmetries are attributable to the gravity-darkened photosphere of the rapidly rotating host star and have been reported for a handful of other transiting systems with hot host stars, including Kepler-13A \citep{barnes2011}, Kepler-462 \citep{ahlers2015}, HAT-P-70 \citep{zhou2019}, and MASCARA-4 \citep{ahlers2020b}. A detailed, dedicated analysis of the gravity-darkened transit of KELT-9b in \tess\ photometry is described in \citet{ahlers2020}. In our phase-curve analysis, we trimmed the transits from the light curve prior to fitting.

During normal spacecraft operation, the photometric observations during each Sector are interrupted by scheduled momentum dumps. These events can cause discontinuities in flux and, in some cases, short-term photometric variability before and/or after. These instrumental artifacts are not fully removed by the PDC pipeline and can present difficulties for the systematics detrending during our fits. Following previous work, we split each orbit's time series into segments separated by the momentum dumps. We found that the first segment of each spacecraft orbit exhibits a discernible initial flux ramp. After inspecting the binned residuals from the phase-curve fit to each segment, we removed the transient flux features by trimming the first 0.25~day worth of data from segments 1, 4, and 7 and the first 1.5~days worth of data from segment 9. The outlier-trimmed, median-normalized light curve is shown in the top panel of Figure~\ref{fig:lc}, with the momentum dumps and light-curve segments indicated in blue and the trimmed points marked in red and green. The final light curve contains 17,489 data points.

\subsection{Phase-curve Model}\label{subsec:model}

Our phase-curve fitting was done using the Python-based ExoTEP pipeline \citep[e.g.,][]{benneke2019,shporer2019,wong2019}. The secondary eclipse light curve $\lambda_{e}(t)$ is modeled using \texttt{batman} \citep{kreidberg2015} and is defined to be zero at mid-eclipse and unity out of eclipse. The phase-curve variation is modeled as an $n$th-order Fourier series at the orbital phase ($\phi\equiv (t-T_0)/P$, where $T_0$ is the transit time, and $P$ is the orbital period) and is split into two components, one describing the photometric signal from the host star and another containing terms attributed to the planet's brightness variation:
\begin{gather}
\label{planet}\psi_{p}(t) = \bar{f_{p}} + B_1 \cos(2\pi\phi+\delta),\\
\label{star}\psi_{*}(t) = 1+ \sum\limits_{k=1}^{n}A_k\sin(2\pi k\phi)+\sum\limits_{k=2}^{n}B_k\cos(2\pi k\phi).
\end{gather}
The planet's atmospheric brightness modulation around its mean value $\bar{f_{p}}$ has a semiamplitude $B_{1}$ and a phase shift $\delta$. It follows that the secondary eclipse depth (i.e., the planet's dayside flux at superior conjunction) is precisely the value of $\psi_{p}(t)$ at phase $\pi$, while the nightside flux of the planet's atmosphere is calculated at phase $0$: $D_{d}=\bar{f_{p}}+B_1\cos(\pi+\delta)$ and $D_{n}=\bar{f_{p}}+B_1\cos(\delta)$. 

In general, the brightness modulation at visible wavelengths is a combination of contributions from reflected and emitted light. While the thermal emission component is well described by a simple sinusoidal function (see, for example, the numerous Spitzer phase-curve studies; e.g., \citealt{wong2016}, \citealt{beatty2019}), the reflected component can take on a more complicated form. In the case of Lambertian scattering, the brightness modulation contains an additional second-order term at the first harmonic of the orbital phase \citep[e.g.,][]{faigler2015}. However, as discussed in Section~\ref{subsec:temp}, the geometric albedo of KELT-9b is consistent with zero, indicating a negligible reflected component in the \tess\ phase curve. Therefore, the model in Equation~\eqref{planet} is expected to provided an accurate description of the atmospheric brightness modulation.

All remaining photometric variability is associated with the host star. Out of these terms, the first harmonic of the cosine $B_{2}$ is attributable to the leading term of the photometric variation arising from ellipsoidal distortion of the host star \citep[e.g.,][]{morris1985,morris1993,pfahl2008}. The second-order term of the ellipsoidal distortion component is at the fundamental of the orbital phase; however, the predicted amplitude of this term is more than an order of magnitude smaller than the leading-order term and well below the typical error bars on the sinusoidal amplitudes from our fits.

We included the fundamental of the sine term in the star's flux model. A signal at this frequency and phase can arise from Doppler boosting \citep[e.g., the \tess\ phase curve of WASP-18;][]{shporer2019}, which stems from periodic blue- and red-shifting of the stellar spectrum, as well as modulations in the photon emission rate in the observer's direction due to the radial velocity of the host star induced by the orbiting planet \citep[e.g.,][]{shakura1987,loeb2003,shporer2010}. In a free, unconstrained fit, the amplitude of this term ($A_{1}$) is degenerate with a phase shift in the planet's brightness modulation in the combined phase curve. To address this degeneracy while still including the term in the phase-curve modeling, we computed the predicted strength of the Doppler boosting signal for the KELT-9 system using the formulation in \citet{esteves2013} and obtained $2.1\pm0.3$~ppm, which we used as a Gaussian prior on $A_{1}$ in the joint fits presented in this paper.

The combined normalized astrophysical phase curve and secondary eclipse model is as follows:
\begin{equation}
    \psi(t) = \frac{\psi_{*}(t)+\lambda_e(t)\psi_{p}(t)}{1+\bar{f_p}}.
\end{equation}
We modeled any remaining long-term photometric variability in the PDC light curves due to residual uncorrected systematics or stellar variability using generalized polynomials in time:
\begin{equation}\label{systematics}
    S_N^{\lbrace i\rbrace}(t) = \sum\limits_{j=0}^{N}c_j^{\lbrace i\rbrace}(t-t_0)^j.
\end{equation}
Here $t_0$ is the time of the first exposure in segment $i$, and $N$ is the order of the detrending polynomial. We optimized the polynomial order of the detrending polynomial for each segment by carrying out individual segment phase-curve fits and considered both the Akaike information criterion ($\mathrm{AIC}\equiv 2\gamma -2 \log L$) and the BIC ($\mathrm{BIC}\equiv \gamma\log m -2 \log L$); here $\gamma$ is the number of free parameters in the fit, $m$ is the number of data points in the segment, and $L$ is the maximum log-likelihood. For the 10 data segments, minimizing the AIC and BIC yielded the same optimal polynomial orders: 2, 1, 1, 0, 0, 1, 4, 4, 1, and 3.

The combined phase-curve and systematics model of KELT-9 is
\begin{equation}\label{fita}
    f(t) = S_N^{\lbrace i\rbrace}(t)|_{i=1-6}\times\psi(t).
\end{equation}
The systematics-removed light-curve segments are shown in the bottom panel of Figure~\ref{fig:lc}.

\begin{figure*}
\includegraphics[width=\linewidth]{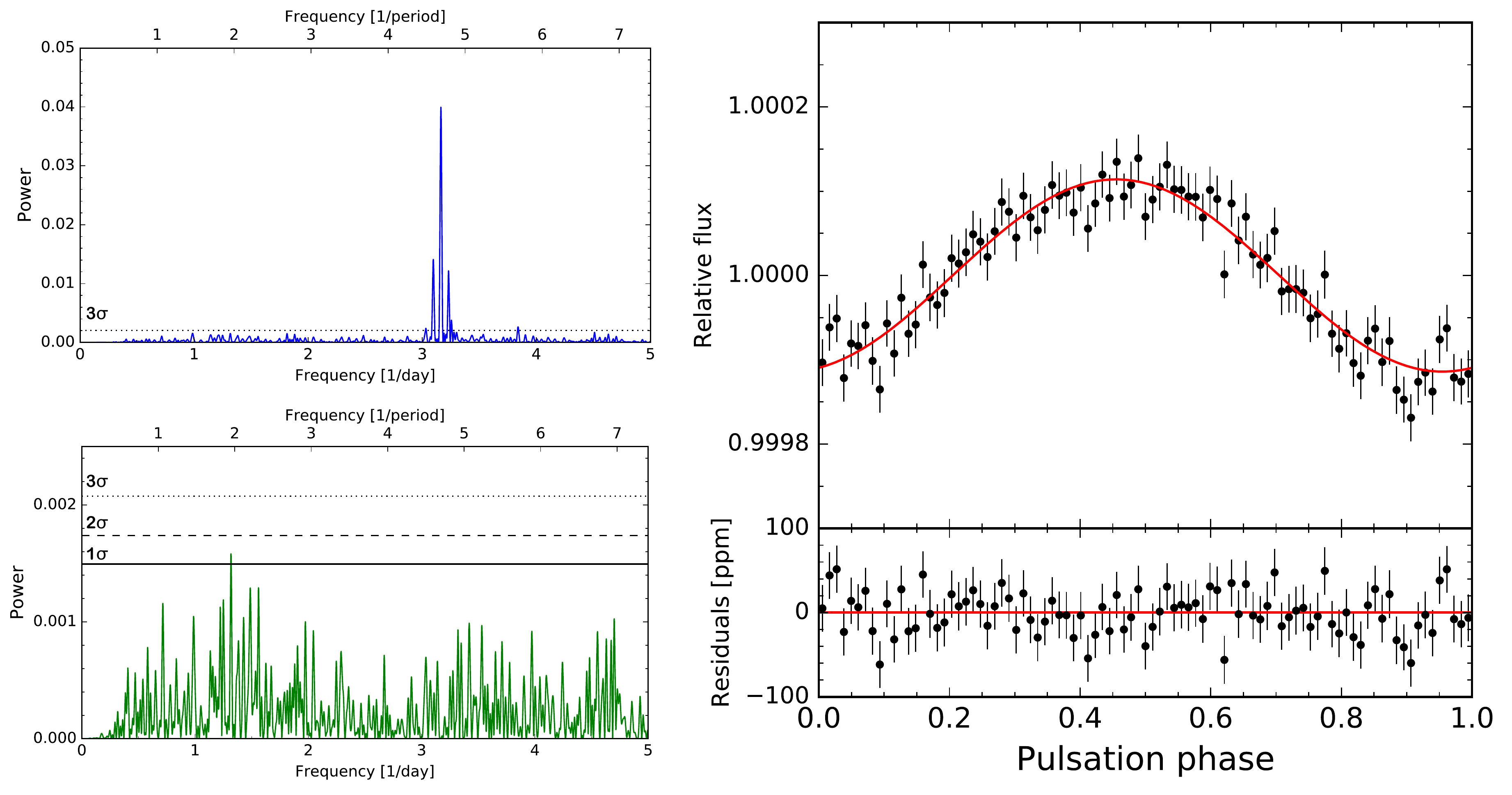}
\singlespace
\caption{Top left: Lomb--Scargle periodogram of the residual array from the joint phase-curve fit without accounting for the stellar pulsations of KELT-9 (fit A). The frequencies are given in units of inverse days, as well as relative to the orbital period of KELT-9b. The strong peak in the power spectrum corresponds to a characteristic pulsation period of 7.59 hr. Bottom left: same as top panel but for the phase-curve fit that includes stellar pulsation modeling (fit B). Horizontal lines indicate statistical significance thresholds. Note the difference in vertical axis scale between the two plots. Right: systematics-corrected light curve and corresponding residuals, phase-folded at the stellar pulsation period, with all other phase-curve signals removed; these data are binned in 5~minute intervals.}
\label{fig:power}
\end{figure*}

\subsection{Stellar Pulsations}\label{subsec:pulsation}

We detected a separate periodic variation in the \tess\ photometry with a frequency unrelated to the orbital period of KELT-9b. The top left panel of Figure~\ref{fig:power} shows the Lomb--Scargle periodogram of the residuals from a joint light-curve fit using the model in Equation~\eqref{fita}. A strong signal with a characteristic period of roughly 0.316~day is evident. This additional periodicity was also detected in the Spitzer 4.5~$\mu$m light curve \citep{mansfield2019}.

We obtained light curves produced by the Quick Look Pipeline \citep{huang2018} for all nearby sources brighter than 13th magnitude in the \tess\ band that lie within $6'$ of KELT-9. When inspecting the individual periodograms, we did not find any significant periodicity at the frequency seen in the KELT-9 light curve. We therefore attribute this periodic signal to stellar pulsations on KELT-9. In our light-curve analysis, we addressed this contribution to the overall photometric variability by simultaneously fitting for the stellar pulsation signal in our joint phase-curve analysis. The stellar pulsations were modeled by a simple sinusoid function with three additional free parameters,
\begin{equation}\label{pulse}
    \Theta(t) = 1 + \alpha\sin\left(2\pi\xi\right) + \beta\cos\left(2\pi\xi\right),
\end{equation}
where $\xi\equiv(t-T_{0})/\Pi$ is the phase given a stellar pulsation period $\Pi$, and $\alpha$ and $\beta$ are the semiamplitudes of the Fourier components of the pulsation signal. This stellar pulsation model is multiplied to the host star's photometric modulation described in Equation~\eqref{star} to produce the corrected astrophysical phase curve and secondary eclipse model:
\begin{equation}\label{corrected}
    \psi'(t) = \frac{\psi_{*}(t)\Theta(t)+\lambda_e(t)\psi_{p}(t)}{1+\bar{f_p}}.
\end{equation}
The full phase-curve model used in our light-curve fits, incorporating both stellar pulsations and systematics, is as follows:
\begin{equation}\label{fitb}
    f'(t) = S_N^{\lbrace i\rbrace}(t)|_{i=1-6}\times\psi'(t).
\end{equation}

We have assumed that the stellar pulsations can be modeled as a simple sinusoid. In general, the shape of the photometric modulation induced by pulsations need not follow such a simple functional form and may consist of additional harmonics and/or other characteristic frequencies. However, we found that including just the fundamental at the stellar pulsation period removed all significant signals in the periodogram of the resultant residuals, as illustrated in the bottom left panel of Figure~\ref{fig:power}. As an independent check, we experimented with adding higher-order terms at the stellar pulsation period or sinusoidal terms with different periods and found no improvement to the AIC or BIC.

\subsection{Model Fitting}\label{subsec:fit}

As discussed previously, we removed the transits from the \tess\ light curve prior to fitting. To constrain the shape of the secondary eclipse, we placed Gaussian priors on the impact parameter $b$, scaled semi-major axis $a/R_{*}$, and planet--star radius ratio $R_{p}/R_{*}$ using the results from a dedicated analysis of the gravity-darkened \tess\ transits \citep{ahlers2020}: $b = 0.14\pm0.04$, $a/R_{*}=3.36\pm0.14$, and $R_{p}/R_{*}=0.0791\pm0.0018$. We also used the transit ephemeris derived from the same analysis as the priors: $T_{0}=2{,}458{,}683.4449\pm0.0003$ BJD$_{\mathrm{TDB}}$ and $P=1.4811235\pm0.0000011$ days.  

We carried out an ensemble of joint fits of all 10 data segments from Sector 14 and 15 \tess\ photometry, varying the number of phase-curve terms included in the astrophysical model. Some of these fits are listed and numbered in Table~\ref{tab:comparison}, with the variables included as free parameters in each instance indicated by check marks. Phase-curve harmonics up to $k=3$ were considered in the astrophysical phase-curve model. In order to examine whether accounting for the stellar pulsations has a notable effect on the phase-curve results, we carried out a joint fit using the phase-curve model in Equation~\eqref{fita}, which does not account for stellar pulsations (\#1; fit A), as well as one that simultaneously fit for the phase curve and stellar pulsations (\#2; fit B), following Equations~\eqref{pulse}--\eqref{fitb}. We also checked for eccentricity in the system via a joint fit that allowed the orbital eccentricity $e$ and argument of periastron $\omega$ to vary freely (\#3). The remaining joint fits (\#4--\#7) contained subsets of the phase-curve terms in fit B, thereby probing the statistical significance of the various detected components independent of the formal uncertainties on the associated amplitudes derived from the fits. 

\begin{deluxetable*}{lcccccccccc}
\tablewidth{0pc}
\tabletypesize{\scriptsize}
\tablecaption{
    AIC/BIC Comparison of Joint Fits
    \label{tab:comparison}
}
\tablehead{\colhead{Fit Number} & 
     \colhead{$A_{1}$} &
    \colhead{$B_{1}$} &
     \colhead{$\delta$} &
      \colhead{$A_{2}$} &
       \colhead{$B_{2}$} &
        \colhead{$A_{3}$,$B_{3}$} &
       \colhead{$\Pi$,$\alpha$,$\beta$\tablenotemark{a}} &
        \colhead{$e$,$\omega$\tablenotemark{a}} &
       \colhead{$\Delta$AIC\tablenotemark{b}} &
       \colhead{$\Delta$BIC\tablenotemark{b}} 
}
\startdata
1 (fit A) & $\checkmark$ & $\checkmark$ & $\checkmark$ & $\checkmark$ & $\checkmark$ & $\checkmark$ & --- & --- & 730 & 710 \\
2 (fit B) & $\checkmark$ & $\checkmark$ & $\checkmark$ & $\checkmark$ & $\checkmark$ & $\checkmark$ & $\checkmark$ & --- & --- & --- \\
3 & $\checkmark$ & $\checkmark$ & $\checkmark$ & $\checkmark$ & $\checkmark$ & $\checkmark$ & $\checkmark$ & $\checkmark$ & 4.7 & 20 \\
4 & $\checkmark$ & $\checkmark$ & $\checkmark$ & $\checkmark$ & $\checkmark$ & --- & $\checkmark$ & --- & 5.1 & $-$10 \\
5 & $\checkmark$ & $\checkmark$ & $\checkmark$ & --- & $\checkmark$ & --- & $\checkmark$ & --- & 90 & 51 \\
6 & $\checkmark$ & $\checkmark$ & --- & $\checkmark$ & $\checkmark$ & --- & $\checkmark$ & --- & 45 & 8.1 \\
7 & $\checkmark$ & $\checkmark$ & $\checkmark$ & --- & --- & --- & $\checkmark$ & --- & 93 & 47 
\enddata
\textbf{Notes.}
\vspace{-0.2cm}\tablenotetext{a}{The three variables $(\Pi,\alpha,\beta)$ are used to model the stellar pulsations. The free eccentricity fits include the orbital eccentricity $e$ and the argument of periastron $\omega$ as fit parameters.}
\vspace{-0.2cm}\tablenotetext{b}{All AIC and BIC values are given relative to fit B, which is the primary analysis of this paper.}
\end{deluxetable*}



The ExoTEP pipeline utilizes the affine-invariant Markov Chain Monte Carlo (MCMC) ensemble sampler \texttt{emcee} \citep{emcee} to simultaneously calculate the posterior distributions of all free parameters. In addition to the astrophysical and systematics detrending parameters, we defined a uniform per-point uncertainty $\sigma_{i}$ for each of the 10 segments to ensure that the reduced $\chi^2$ value is unity and self-consistently generate realistic uncertainties on the fitted parameters. The number of walkers was set to four times the number of free parameters. The length of each chain was 50,000 steps, and we used only the last 40\% of each chain for calculating the posterior distributions. To check for convergence, we applied the Gelman--Rubin test \citep{gelmanrubin} and ensured that the diagnostic value $\hat{R}$ was well below 1.1. 


\begin{deluxetable*}{lllll}
\tablewidth{0pc}
\tabletypesize{\scriptsize}
\tablecaption{
    Results of Joint Fits
    \label{tab:fit}
}
\tablehead{\\ & \multicolumn{2}{c}{\underline{Fit A}\tablenotemark{a}}  & \multicolumn{2}{c}{\underline{Fit B}\tablenotemark{a}} \\
    Parameter &
    \colhead{Value}                     &
    \colhead{Error}  &
    \colhead{Value}                     &
    \colhead{Error}  
}
\startdata
\sidehead{Orbital and System Parameters}
$T_0$ (BJD$_{\mathrm{TDB}}-2{,}458{,}000$)\tablenotemark{b}        & 711.58620 & $_{-0.00024}^{+0.00025}$ &  711.58627 & $_{-0.00024}^{+0.00025}$ \\
$P$ (days)\tablenotemark{b}    & 1.4811236 & $_{-0.0000011}^{+0.0000010}$ & 1.4811235 & 0.0000011\\
$R_{p}/R_{*}$\tablenotemark{b}  & 0.0792 & 0.0018 & 0.0790 & $_{-0.0017}^{+0.0018}$ \\
$b$\tablenotemark{b}  & 0.138 & $_{-0.041}^{+0.037}$  & 0.134 & $_{-0.038}^{+0.040}$ \\
$a/R_{*}$\tablenotemark{b}  & 3.189 & $_{-0.025}^{+0.022}$  & 3.191 & $_{-0.025}^{+0.022}$ \\
\sidehead{Phase-Curve Parameters}
$\bar{f_p}$ (ppm)    & 381 & $_{-12}^{+13}$ & 369 & $_{-13}^{+11}$ \\
$A_1$ (ppm)\tablenotemark{c} & 2.1 & 0.3 & 2.1 & 0.3 \\
$B_1$ (ppm)   & $-$295.7 & $_{-8.1}^{+7.6}$ & $-$283.0 & $_{-7.8}^{+7.9}$\\
$\delta$ (deg)    &  5.5 & $_{-0.9}^{+0.8}$ & 5.2 & 0.9 \\
$A_2$ (ppm)   & $-$32.3 & $_{-4.1}^{+4.2}$ & $-$35.7 & $_{-4.3}^{+4.2}$ \\
$B_2$ (ppm)   & 11.6 & $_{-6.0}^{+6.1}$ & 16.1 & $_{-5.9}^{+6.0}$ \\
$A_3$ (ppm)   & 18.9 & $_{-4.2}^{+4.1}$ & 13.9 & $_{-4.3}^{+4.2}$ \\
$B_3$ (ppm)   & $-$13.8 & $_{-6.6}^{+6.8}$ & $-$3.0 & $_{-6.2}^{+6.4    }$ \\
\sidehead{Stellar Pulsation Parameters}
$\Pi$ (hr)  & $\dots$ & $\dots$ & 7.58695 & 0.00091 \\
$\alpha$ (ppm)   & $\dots$ & $\dots$ & 31.9 & $_{-4.3}^{+4.2}$\\
$\beta$ (ppm)   & $\dots$ & $\dots$ & $-$109.5 &  $_{-4.3}^{+4.2}$ \\
\sidehead{Derived Parameters} 
Secondary eclipse depth, $D_{d}$ (ppm)  & 676 & 15 & 650 & $_{-15}^{+14}$\\
Nightside flux, $D_{n}$ (ppm)  & 87 & $_{-14}^{+15}$ & 87 & 14 \\
Inclination, $i$ (deg) & 87.53 & $_{-0.69}^{+0.74}$ & 87.60 & $_{-0.73}^{+0.69}$ \\
Dayside brightness temperature, $T_{B,p,\mathrm{day}}$ (K)\tablenotemark{d} & 4640 & 100 & 4600 & 100\\
Nightside brightness temperature, $T_{B,p,\mathrm{night}}$ (K) & 3040 & 100 & 3040 & 100 \\
Bond albedo, $A_{B}$\tablenotemark{d} & 0.17 & $^{+0.12}_{-0.11}$ & 0.19 & $^{+0.12}_{-0.11}$ \\
Recirculation efficiency, $\epsilon$ & 0.38 & 0.05 & 0.39 & 0.05
\enddata
\textbf{Notes.}
\vspace{-0.2cm}\tablenotetext{a}{Fit A: joint phase-curve fit without accounting for stellar pulsations. Fit B (primary analysis of the paper): simultaneous fit of phase curve and stellar pulsations.}
\vspace{-0.2cm}\tablenotetext{b}{Constrained by priors based on values from \citet{ahlers2020} (see Section~\ref{subsec:fit}).}
\vspace{-0.2cm}\tablenotetext{c}{Constrained by priors computed following \citet{esteves2013}: $2.1\pm0.3$~ppm.}
\vspace{-0.2cm}\tablenotetext{d}{Assuming zero geometric albedo; see Section~\ref{subsec:temp}.}
\end{deluxetable*}

\section{Results}
\label{sec:res}

As shown in Table~\ref{tab:comparison}, fit B, which includes phase-curve terms up to third order and stellar pulsation modeling, yields the lowest AIC value among the ensemble of joint fits we carried out. In the following discussion, we utilize the values from fit B as the primary results of the paper. Table~\ref{tab:fit} lists the medians and $1\sigma$ uncertainties for all free parameters in fits A and B. From the stellar pulsation modeling in fit B, we obtained a characteristic pulsation period of $7.58695\pm0.00091$~hr and a combined peak-to-peak amplitude of $2\times\sqrt{\alpha^2+\beta^2}=228.1\pm8.5$~ppm. The right panel of Figure~\ref{fig:power} shows the binned light curve phase-folded at the pulsation period, with all other systematics and astrophysical signals removed.

\begin{figure}
\includegraphics[width=\linewidth]{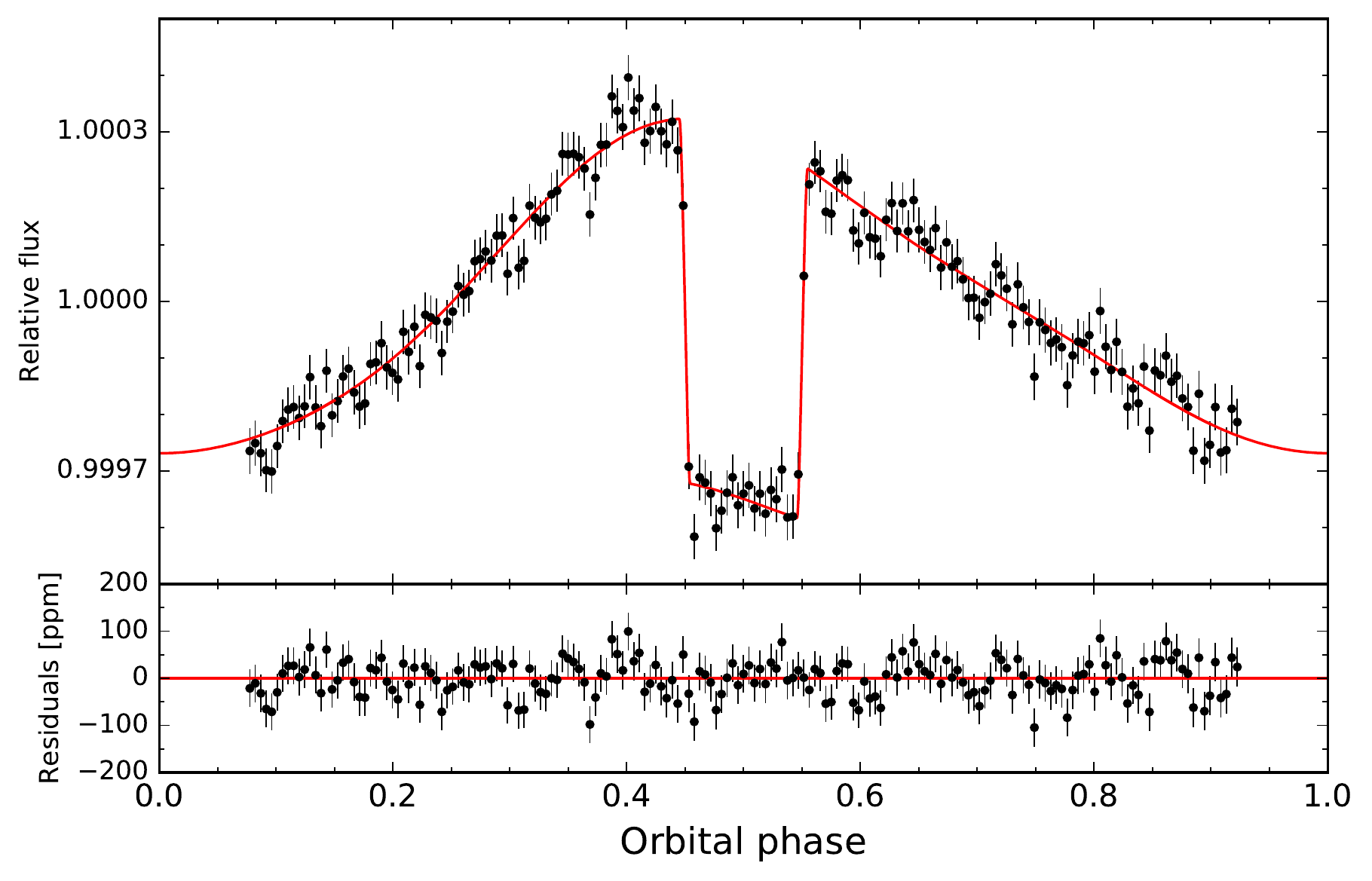}
\singlespace
\caption{Top panel: phase-folded light curve of KELT-9 after correcting for the stellar pulsation signal and long-term trends, binned in 10 minute intervals (black points). The best-fit full phase-curve model from our joint analysis is shown by the red curve. Bottom panel: corresponding residuals from the best-fit model.}
\label{fig:fit}
\end{figure}

For all free and derived parameters common to fits A and B, the measured values agree with each other to within $1.3\sigma$. This comparison demonstrates that the presence of the stellar pulsations, as well as our efforts to address it in the phase-curve fit using a simple sinusoidal model, do not induce any significant biases in the results. The phase-folded, systematics- and stellar pulsation--corrected light curve is shown in Figure~\ref{fig:fit} along with the best-fit full phase-curve model and corresponding residuals.

We detected a strong atmospheric brightness modulation signal with a semiamplitude of $283.0^{+7.8}_{-7.9}$~ppm and a small but statistically significant phase shift of $\delta=5\overset{\circ}{.}2\pm0\overset{\circ}{.}9$; fixing the phase shift to zero incurs relative penalties of $\Delta\mathrm{AIC}=40$ and $\Delta\mathrm{BIC}=18$. As mentioned in Section~\ref{subsec:model}, we placed priors on the expected Doppler boosting signal, which occurs at the same harmonic as the atmospheric brightness modulation. To assess how much of a difference this inclusion makes, we carried out a separate fit with the assumed $A_1=2.1\pm0.3$ Doppler boosting signal removed, obtaining a phase shift of $\delta=5\overset{\circ}{.}6\pm0\overset{\circ}{.}9$, which is statistically identical to the results from our fits including Doppler boosting. Therefore, the effect of Doppler boosting does not significantly affect the phase curve we measure. 

\begin{figure}
\includegraphics[width=\linewidth]{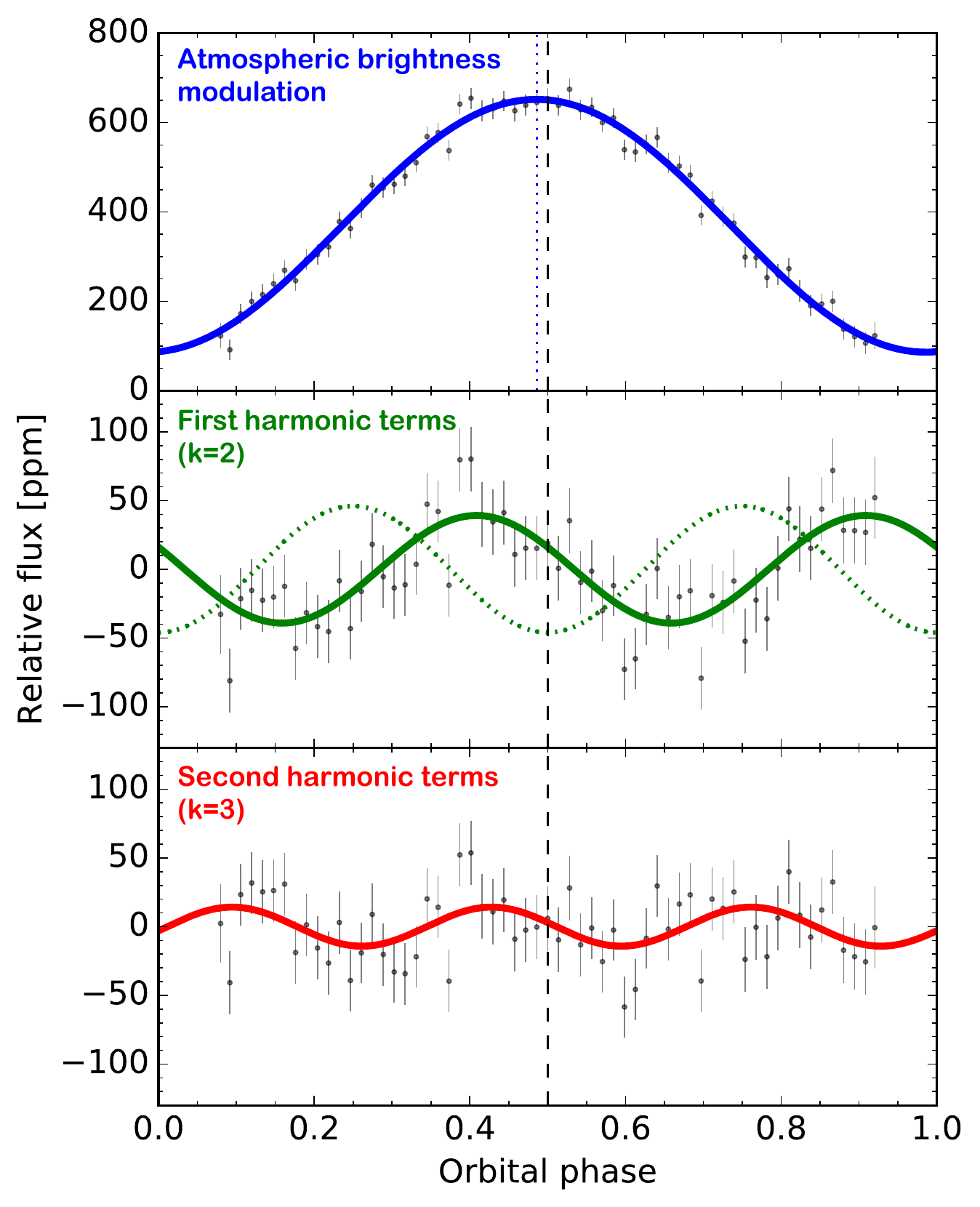}
\singlespace
\caption{Schematic of the phase-curve components measured in our phase-curve analysis of KELT-9. In each panel, the phase-folded light curve is shown with all but the corresponding phase-curve component removed, binned in 30 minute intervals (gray points). The top panel shows the planet's atmospheric brightness modulation signal with the secondary eclipse removed. The blue vertical dotted line marks the phase of maximum brightness. The mid-eclipse phase (0.5) is indicated by the vertical dashed line. In the middle panel, the solid curve displays the measured combined variation at the first harmonic of the orbital phase; the dotted curve represents the nominal phase alignment expected from ellipsoidal distortion of the host star. The bottom panel is a plot of the unexpected marginal second harmonic phase-curve component ($k=3$). Note the difference in the Y-axis scale between the top panel and the other two panels.}
\label{fig:comp}
\end{figure}

From this phase shift, we infer a slight eastward offset in the location of KELT-9b's maximum dayside brightness relative to the substellar point. Given the fitted average planet flux $\bar{f_{p}}$ and atmospheric brightness semiamplitude $B_1$, we used our parameterization of the planet's flux in Equation~\eqref{planet} to obtain the secondary eclipse depth and nightside flux (disk-integrated brightness of the planet at mid-transit): $650^{+14}_{-15}$ and $87\pm14$~ppm, respectively. The top panel of Figure~\ref{fig:comp} shows the variation in KELT-9b's disk-integrated brightness with orbital phase.

Given the high signal-to-noise secondary eclipse detection, we searched for orbital eccentricity by allowing $e$ and $\omega$ to vary as the transit ephemeris was constrained by priors derived from the contemporaneous transit fit by \citet{ahlers2020}. The addition of these two parameters is disfavored by the AIC and BIC, and we did not find any significant deviation from a circular orbit: $e<0.007$ at $2\sigma$, $e\cos \omega = 0.00009_{-0.00039}^{+0.00050}$, $e\sin\omega = 0.0000_{-0.0018}^{+0.0025}$.

We measured significant phase-curve coefficients at the first harmonic ($k=2$) of the orbital phase. As mentioned in Section~\ref{subsec:model}, the coefficient of the cosine at this harmonic $B_{2}$ is the expected leading-order contribution from the ellipsoidal distortion of the host star; standard tidal distortion theory does not predict a signal in the associated sine term $A_{2}$. However, in our joint fit, we find a significant $8.5\sigma$ $A_{2}$ value, and fixing it to zero results in large increases in the AIC and BIC (Table~\ref{tab:comparison}). Combining the sine and cosine terms at the first harmonic produces a single sinusoidal modulation with a semiamplitude of $39.6\pm4.5$~ppm ($8.8\sigma$) that comes to maximum at $0.158^{+0.011}_{-0.012}$ in orbital phase ($5.6\pm0.4$~hr) after quadrature. This combined first harmonic modulation is illustrated in isolation in Figure~\ref{fig:comp}. We discuss the various implications of this phase curve signal in Section~\ref{subsec:ellip}.

A marginal phase-curve modulation at the second harmonic ($k=3$) was detected in our joint fits, with a combined semiamplitude of $15.5^{+4.4}_{-4.3}$~ppm. As shown in Table~\ref{tab:comparison}, the inclusion of these terms ($A_{3}$,$B_{3}$) is somewhat favored by the AIC ($\Delta\mathrm{AIC}=-5.1$) and disfavored by the BIC, which incurs a much larger penalty for additional free parameters, especially given the large number of data points ($\Delta\mathrm{BIC}=10$). The bottom panel of Figure~\ref{fig:comp} shows this modulation with all other phase-curve terms removed.


In order to probe for possible temporal variability in the atmosphere of KELT-9b, we carried out joint fits of data from each spacecraft orbit separately (i.e., segments 1--3, 4--6, 7--8, and 9--10). Both the secondary eclipse depths and the semiamplitudes and phase shifts of the atmospheric brightness modulation are self-consistent to within $1.7\sigma$, indicating no evidence for variability across the dataset.

\section{Discussion}
\label{sec:dis}

The phase-curve analysis described in the previous sections provides a detailed look at the atmospheric brightness modulation of KELT-9b across all longitudes. In addition, we have detected a significant signal at the first harmonic of the orbital period. In this section, we interpret these results to derive constraints on the temperature distribution of KELT-9b, place the planet into the broader context of gas giants with measured phase curves, and explore the potential for future intensive atmospheric characterization.

\subsection{Brightness Temperature Distribution}
\label{subsec:temp}

At superior conjunction, the planet--star flux ratio (i.e., secondary eclipse depth) can be expressed as a function of various fundamental properties of the system \citep[e.g.,][]{charbonneau2005}:
\begin{equation}\label{Dd}
    D_d = \left(\frac{R_{p}}{R_{*}}\right)^{2}\frac{\int \tau(\lambda)F_{p,\nu}(\lambda,T_{B,p,\mathrm{day}})d\lambda}{\int \tau(\lambda)F_{*,\nu}(\lambda,T_{\mathrm{eff}})d\lambda} + A_{g}\left(\frac{R_{p}}{a}\right)^2.
\end{equation}
The first term describes the contribution from the planet's thermal emission. Here $R_{p}/R_{*}$ is the planet-star radius ratio, $T_{\mathrm{eff}}$ is the host star's effective temperature, and $T_{B,p,\mathrm{day}}$ is the disk-averaged brightness temperature of the planet's observer-facing hemisphere. Here $F_{p,\nu}(\lambda,T_{p})$ and $F_{*,\nu}(\lambda,T_{\mathrm{eff}})$ are the photon-weighted planetary and stellar fluxes, which are computed from the corresponding unweighted functions (in wavelength units) via the multiplicative conversion factor $\lambda/hc$. These fluxes are integrated across the \tess\ transmission function $\tau(\lambda)$. The second term represents reflected light off the planet's dayside atmosphere, where $A_{g}$ is the geometric albedo in the \tess\ bandpass, and $a$ is the orbital semi-major axis.

To account for the uncertainties in the stellar parameters and propagate them to the brightness temperature estimates, we calculated the integrated stellar flux in the \tess\ bandpass for a grid of \texttt{PHOENIX} models \citep{husser2013} spanning the ranges $T_{\mathrm{eff}}=[8000,12{,}000]$~K, $[\mathrm{M/H}]=[-1.0,+0.5]$, and $\log g=[3.5,4.5]$. To obtain a simple empirical model of the integrated stellar flux, we fit a linear polynomial in ($T_{\mathrm{eff}}$, $\lbrack\mathrm{M/H}\rbrack$, $\log g$) to the calculated values; including higher-order terms did not improve the fit. We computed the posterior distribution of the dayside brightness temperature using a Monte Carlo sampling method: in order to account for uncertainties in the system parameters, we allow them to vary in the MCMC fit while constraining them with Gaussian priors derived from their published values and uncertainties. For the stellar parameters, we use the measurements reported by \citeauthor{gaudi2017} (2017; $T_{\mathrm{eff}}=10{,}170\pm450$~K, $\log g = 4.093\pm0.014$, $\lbrack\mathrm{M/H}\rbrack = -0.03\pm0.020$), while for $R_{p}/R_{*}$ and $a/R_{*}$, we use the values from \citet{gaudi2017} and \citet{mansfield2019}, respectively ($R_{p}/R_{*}=0.08004\pm0.00041$, $a/R_{*}=3.153\pm0.011$), which have significantly smaller uncertainties than the \tess-band measurements published in \citet{ahlers2020}. We assume a blackbody for the planet's emission spectrum. Both direct measurements and atmospheric models demonstrate that hot Jupiters tend to have very low reflectivities \citep[e.g.,][]{hengdemory2013}. Furthermore, the extremely hot dayside of KELT-9b is expected to preclude the formation of any condensate clouds that might enhance the geometric albedo (see Section~\ref{subsec:context}). Here we allow for a range of geometric albedos spanning zero through 0.2. 

For zero geometric albedo, we derive a dayside brightness temperature of $T_{B,p,\mathrm{day}}=4600\pm100$~K. Across the range of geometric albedos we consider, the dayside temperature varies from 4360 to 4600~K. The measured dayside temperature of KELT-9b is comparable to those of mid-K dwarfs. An analogous calculation using the measured nightside flux (and no reflected light) yields a very high nightside temperature of $T_{B,p,\mathrm{night}}=3040\pm100$~K. This means that the nightside of KELT-9b is hotter than the dayside of almost all known exoplanets and comparable to the atmosphere of a mid-M dwarf.

The difference between the dayside and nightside temperatures of a planet reflects the efficiency of heat transport from the dayside to the nightside, e.g., by zonal winds. The more inefficient the heat recirculation is, the larger the expected day-night temperature contrast. Following the simple thermal balance considerations in \citet{cowanagol}, the relationship between irradiation, heat recirculation, and dayside and nightside temperatures can be expressed as
\begin{gather}
    T_{B,p,\mathrm{day}}=T_{*}\sqrt{\frac{R_{*}}{a}}(1-A_{B})^{1/4}\left(\frac{2}{3}-\frac{5}{12}\epsilon\right)^{1/4},\label{tday}\\
    T_{B,p,\mathrm{night}}=T_{*}\sqrt{\frac{R_{*}}{a}}(1-A_{B})^{1/4}\left(\frac{\epsilon}{4}\right)^{1/4},\label{tnight}
\end{gather}
where $A_{B}$ is the Bond albedo (i.e., the fraction of total incident stellar irradiation that is reflected), and $\epsilon$ is a fiducial parameter that indicates the relative efficiency of heat transport, with zero corresponding to no day--night heat recirculation and 1 corresponding to full recirculation. 

By simultaneously fitting to the measured dayside and nightside temperatures in the \tess\ bandpass using Equations~\eqref{tday} and \eqref{tnight} and constraining $T_{*}$ and $a/R_{*}$ with the same Gaussian priors presented above while keeping $\epsilon$ and $A_{B}$ unconstrained, we obtain $A_{B}=0.19^{+0.12}_{-0.11}$ and $\epsilon=0.39\pm0.05$. These results indicate relatively efficient day--night heat transport. All of the quantities derived in this subsection are listed in Table~\ref{tab:fit}. The measured heat recirculation parameter will be placed into context with other planets in Section~\ref{subsec:context}.

\subsubsection{Comparison to \citet{mansfield2019}}\label{subsubsec:mansfield}
\citet{mansfield2019} analyzed the full-orbit Spitzer phase curve of the KELT-9 system, obtained in the 4.5~$\mu$m bandpass. They measured a high signal-to-noise atmospheric brightness modulation and a secondary eclipse depth of $3131\pm62$~ppm. Following an analogous calculation to the one described above, they reported a dayside brightness temperature of $4566^{+140}_{-136}$~K. This temperature is statistically identical to our estimate assuming zero geometric albedo, indicating that the dayside emission spectrum is consistent with a single blackbody and negligible atmospheric reflectivity at visible wavelengths. 

For the dayside, we can simultaneously fit the \tess\ and Spitzer eclipse depths to obtain better constraints on the dayside temperature. We also include the $z'$-band eclipse depth of $1006\pm97$~ppm from the discovery paper \citep{gaudi2017}. This three-point fit yields $T_{\mathrm{day}}=4540\pm90$~K when assuming zero geometric albedo. Allowing the stellar and system parameters to vary within Gaussian priors is crucial in producing a good fit, with the resulting best-fit $T_{\mathrm{eff}}$ and $R_{p}/R_{*}$ median and $1\sigma$ errors being $9880\pm420$~K and $0.08023\pm0.00039$, respectively; both of these posterior distributions are somewhat shifted from the prior distributions ($T_{\mathrm{eff}}=10{,}170\pm450$~K, $R_{p}/R_{*}=0.08004\pm0.00041$), though still consistent at much better than the $0.5\sigma$ level.


The corresponding nightside brightness temperature estimate from the Spitzer 4.5~$\mu$m phase curve is $2556^{+101}_{-97}$~K. This value is significantly lower than the brightness temperature we measured from the \tess-band nightside flux: $3040\pm100$~K. The lower brightness temperature in the infrared indicates that the nightside emission spectrum of KELT-9b deviates significantly from that of a simple blackbody, likely due to H$^-$ opacity (see Section~\ref{subsec:atmo} for more discussion). The lower nightside brightness temperature measured by \citet{mansfield2019} also translates to poorer day--night heat recirculation at the pressure levels probed by the 4.5~$\mu$m bandpass. Plugging the Spitzer-derived dayside and nightside temperatures into Equations~\eqref{tday} and \eqref{tnight}, we obtained $A_{B,4.5}=0.28^{+0.13}_{-0.15}$ and $\epsilon_{4.5}=0.23\pm0.04$. While the poorly constrained Bond albedo values are consistent between the two wavelengths, the recirculation efficiencies differ by $2.5\sigma$.

The last point of comparison between these two studies is the measured phase offset in the atmospheric brightness modulation. While the \tess-band offset is $5\overset{\circ}{.}2\pm0\overset{\circ}{.}9$, \citet{mansfield2019} obtained $18\overset{\circ}{.}7^{+2\overset{\circ}{.}1}_{-2\overset{\circ}{.}3}$ at 4.5~$\mu$m, more than $5\sigma$ larger. Within the simplistic view that a larger hot-spot offset is indicative of more efficient day--night heat recirculation, the larger phase-curve offset measured in the infrared is at odds with the larger day--night temperature contrast.

However, as discussed in the Introduction, other recent phase-curve observations of highly irradiated exoplanets have blurred the correlation between insolation, heat recirculation, and phase-curve offsets. More generally, it is expected that longitudinal gradients in atmospheric composition can lead to substantially different temperature--pressure profile shapes between the dayside and nightside hemispheres, which subsequently modulates the photospheric pressure levels across the planet. Likewise, the interplay between irradiation, composition, three-dimensional atmospheric dynamics, and local temperature--pressure profiles may not be adequately captured by simple one-dimensional thermal balance models, such as the one we used to calculate the recirculation efficiencies. Lastly, atmospheric modeling has demonstrated the possibility of temporal variability in the atmospheric dynamics of hot Jupiters, which can be manifested by time-varying hot-spot offsets across different epochs \cite[e.g.,][]{rogers2017,komacek2020}. We address some of these complexities when discussing the atmospheric modeling of KELT-9b in Section~\ref{subsec:atmo}.

\subsection{First and Second Harmonic Terms}
\label{subsec:ellip}

In binary systems, the orbiting companion raises a tidal bulge on the surface of the host star, which incurs periodic variations in the star's sky-projected area. The observer-facing profile of the host star is expected to come to maximum at quadrature, producing a photometric modulation with a leading term at the first harmonic of the cosine. The amplitude of this variation can be related to fundamental properties of the system via the theoretical formalism described in \citet{kopal1959}. Using the notation of \citet{morris1985} and \citet{morris1993}, we can express the dominant amplitude $B_{2}$ as
\begin{equation}
    B_{2} = -Z_{\mathrm{ellip}}\frac{M_{p}}{M_{*}}\left(\frac{R_*}{a}\right)^3\sin^2{i},
\end{equation}
where $M_{p}/M_{*}$ is the planet--star mass ratio, and the prefactor $Z_{\mathrm{ellip}}$ depends on the star's linear limb- and gravity- darkening coefficients and is of order unity. The planet also experiences ellipsoidal distortion, but the amplitude is smaller than the stellar signal by a factor of at least $(R_{p}/R_{*})^3\bar{f}_{p}/(M_{p}/M_{*})^2\sim 0.5\%$ and can thus be safely ignored.

Using limb- and gravity-darkening coefficient values computed by \citet{claret2017} in the \tess\ bandpass for the nearest available combination of stellar parameters ($\log T_{\mathrm{eff}}=3.978$, $\log g=4.00$, $\mathrm{[Fe/H]}=0.10$, $u_{1}=0.3494$, $\gamma_{1}=0.4355$), we calculate a predicted ellipsoidal distortion amplitude of $44\pm6$~ppm. Beyond the leading-order term, there are smaller contributions at the second harmonic of the cosine and the fundamental of the orbital period (i.e., $\cos(6\pi\phi)$ and $\cos(2\pi\phi)$). Using the formulas in \citet{wong2019kepler}, we compute these expected amplitudes to be $2.7\pm0.4$ and $1.6\pm0.2$~ppm, respectively.

Looking back to the results from our phase-curve fit, we find that while the total amplitude of the photometric variation at the first harmonic ($39.6\pm4.5$~ppm) is consistent with the predicted amplitude of $44\pm6$~ppm, there is a very significant phase offset in the modulation relative to the expectation. In Figure~\ref{fig:comp}, we plot the measured first harmonic modulation (solid curve) alongside the predicted signal at the expected phasing (dotted curve). 

We also measured a phase-curve variation at the second harmonic with an amplitude of $15.5^{+4.4}_{-4.3}$~ppm, and once again, there are significant discrepancies in both amplitude and phasing between the measured signal and the predicted second-order term of the ellipsoidal distortion. Given the relatively low significance of the second harmonic detection, the measured signal may be due to residual short-term instrumental systematics or other noise in the photometry. Looking at Figure~\ref{fig:fit}, we can see that the phase-folded residuals from the joint fit contain some traces of time-correlated noise. When the KELT-9 system is revisited by \tess\ during the Extended Mission, we will obtain significantly more photometry, from which more robust constraints can be placed on the signal at the second harmonic. We note that unexpected second harmonic phase-curve signals have been detected in other systems, including HAT-P-7 \citep{armstrong2016} and Kepler-13A/KOI-13 \citep{esteves2013, shporer2014}. 

Mismatches between the measured and expected ellipsoidal distortion signals have been previously reported in other systems with early-type host stars, including KOI-54 \citep{burkart2012}, KOI-74 \citep{rowe2010, vankerkwijk2010, ehrenreich2011, bloemen2012}, and KOI-964 \citep{carter2011,wong2019kepler}. These discrepancies might arise due to an oversimplification of the tidal dynamics of hot stars with largely radiative envelopes, where interactions between the dynamical tide and the gravitational influence of the orbiting companion can yield large steady-state deviations in both amplitude and relative phase from the equilibrium tide predictions \citep{pfahl2008}. 
Modeling of the tidal response of KOI-54 ($M=2.32\pm0.10\,M_{\Sun}$, $R = 2.19\pm0.03\,R_{\Sun}$) --- an A-type star similar to KELT-9 ($M=2.32\pm0.16\,M_{\Sun}$, $R = 2.418\pm0.058\,R_{\Sun}$; \citealt{borsa2019}) --- revealed phase shifts in the resultant ellipsoidal distortion modulation of up to a quarter of an orbit, depending on the proximity of the orbital period to resonances and the damping of the stellar oscillation modes \citep{burkart2012}.


\subsection{Consequences of Rapid Stellar Rotation and KELT-9b's Polar Orbit}\label{subsec:irradiance}

The host star's rapid rotation, combined with the peculiar near-polar orbit of KELT-9b, may also explain the phase shift in the measured first harmonic modulation. \citet{ahlers2020} modeled the gravity-darkened transits of KELT-9b from the \tess\ light curve and simultaneously resolved the three-dimensional stellar spin axis and the orbital misalignment. They measured a sky-projected stellar obliquity of $\lambda=-88^{\circ}\pm15^{\circ}$, a stellar inclination of $\psi=-52^{\circ} (+8^{\circ},-7^{\circ})$, and a spin-orbit misalignment of $\varphi=87^{\circ} (+10^{\circ},-11^{\circ})$. When combined with the orbital inclination measurement of $i=87\overset{\circ}{.}2\pm0\overset{\circ}{.}4$, this means that KELT-9b passes almost perfectly over the stellar poles and that the spin axis of the host star lies in a plane that is nearly perpendicular to the sky plane. Furthermore, given the stellar inclination, KELT-9b passes over the stellar pole roughly one-sixth of an orbit prior to mid-transit and mid-eclipse.

KELT-9 has a rapid spin, with a measured sky-projected rotational velocity of $v\sin i_{*}=114\pm1.3$~km~s$^{-1}$ \citep{gaudi2017} and a stellar rotation period of $16^{+3}_{-4}$~hr; the rapid rotation also generates a substantial stellar oblateness of $0.089\pm0.017$ \citep{ahlers2020}. Due to gravity darkening, the star is brightest at the poles, and as the planet orbits around the host star, the nonhomogeneous stellar effective temperature distribution alters the irradiation of the planet. Meanwhile, the stellar oblateness changes the size of the projected stellar disk that the planet's dayside sees across the orbit. These two effects combine constructively to produce an irradiance modulation that comes to maximum twice per orbit, when the planet passes near the stellar pole. The corresponding periodic photometric modulation stemming from these processes contributes an additional signal in the planet's flux at the first harmonic of the orbital period.

\begin{figure}
\includegraphics[width=\linewidth]{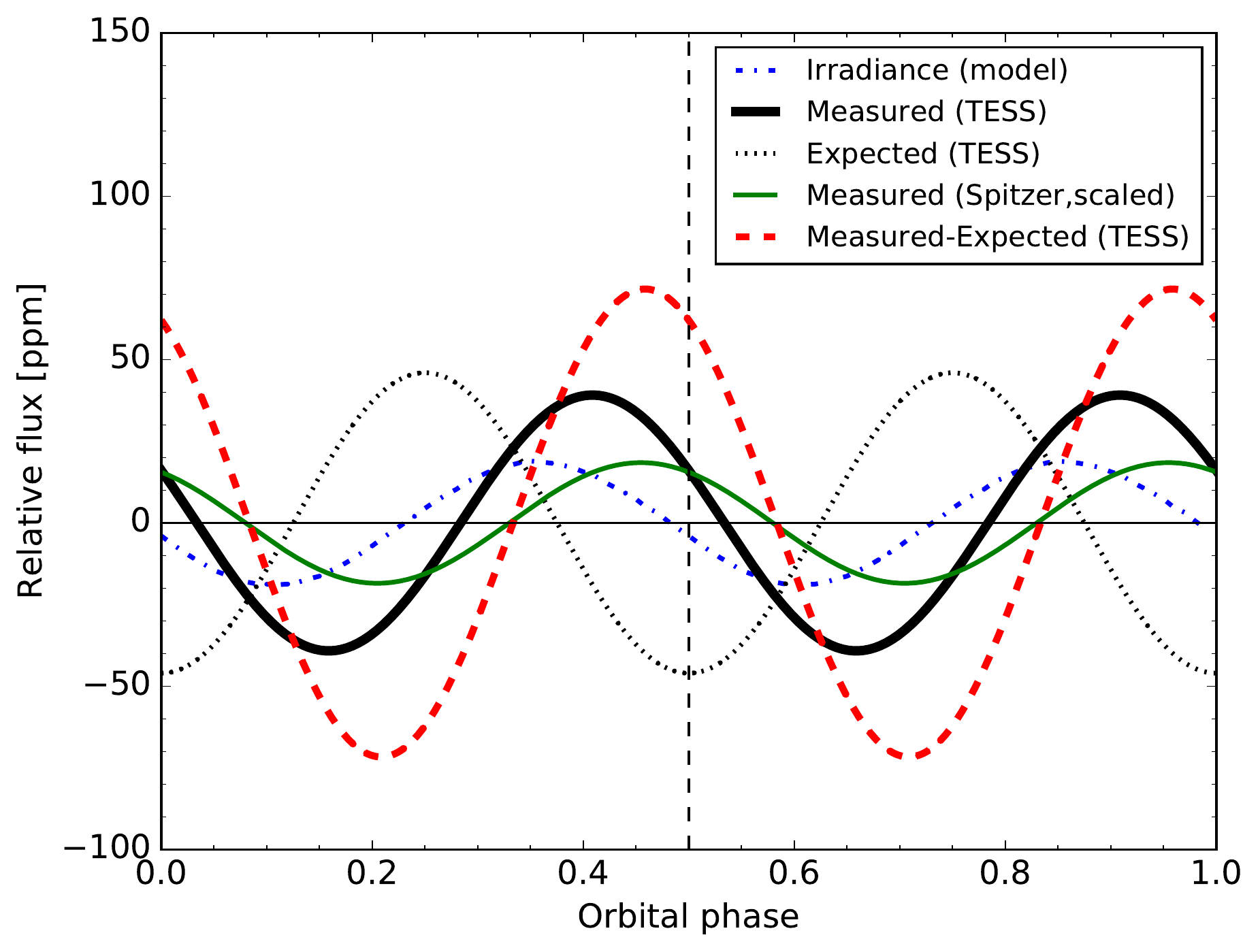}
\singlespace
\caption{Comparison plot of the first harmonic phase-curve component detected in the \tess\ light-curve analysis (solid black curve) with various predicted and measured signals. The thin dotted black curve shows the predicted ellipsoidal distortion modulation, following the physical formalism of equilibrium tide theory (Section~\ref{subsec:ellip}). The blue dotted--dashed curve denotes the modeled photometric signal stemming from the combined effects of the nonuniform stellar brightness distribution and rotationally induced oblateness on the irradiance received by KELT-9b's dayside hemisphere. The thin solid green curve plots the measured first harmonic signal from the Spitzer 4.5~$\mu$m phase curve \citep{mansfield2019}, scaled by the ratio between the relative dayside brightnesses of KELT-9b in the two bandpasses. The dashed red curve shows the measured first harmonic phase-curve signal, subtracted by the expected ellipsoidal distortion modulation. The agreement in the phase alignment of this corrected curve and the Spitzer measurement supports the hypothesis that the large phase shift in the first harmonic component is primarily due to the additional modulation in the planet's thermal emission due to the time-varying irradiance across its orbit.} 
\label{fig:irrad}
\end{figure}

From the gravity darkening and oblateness modeling in \citet{ahlers2020}, we produced an effective temperature map of the host star and generated the total irradiance $I$ experienced by KELT-9b as a function of orbital phase, using the model described in \citet{ahlers2016}. The equilibrium dayside temperature of KELT-9b is related to the incident irradiance via $T_{\mathrm{day}}\propto I^{1/4}$, and we computed the relative change in thermal emission from the planet associated with the varying dayside temperature (see the numerator in the first term of Equation~\eqref{Dd}). After scaling the resultant modulation by the secondary eclipse depth (i.e., dayside flux relative to the star's brightness), we obtained the predicted phase-curve contribution from the planet's time-varying irradiance, plotted in Figure~\ref{fig:irrad}.

From the figure, it is evident that the phasing of this irradiance signal is significantly offset from the expected ellipsoidal distortion modulation. Taking into account the uncertainties in the stellar spin axis solution, the pre-eclipse irradiance maximum occurs at an orbital phase of $0.36\pm0.03$. Notably, the alignment of the irradiance signal is consistent at $1.5\sigma$ with that of the measured first harmonic phase-curve component from the \tess\ light curve, which has a pre-eclipse maximum at an orbital phase of $0.408^{+0.011}_{-0.012}$. 



For additional observational perspectives on the role of periodic atmospheric heating modulations, we turn to the published full-orbit 4.5~$\mu$m thermal phase curve. \citet{mansfield2019} detected a significant phase-curve signal at the first harmonic of the orbital period with a semiamplitude of $89\pm22$~ppm and a pre-eclipse maximum that occurs at an orbital phase of $0.455\pm0.022$. In Figure~\ref{fig:irrad}, we plot this measured first harmonic modulation, with the amplitude scaled by the ratio of KELT-9b's measured relative dayside brightnesses in the \tess\ and Spitzer 4.5~$\mu$m bandpasses: 650/3131. The adjusted amplitude is statistically identical to that of the irradiation model, while the phase alignment of the maxima is consistent with the measurement from the \tess\ phase curve at the $1.9\sigma$ level.

Lastly, if we assume that the tidal response of the star occurs as predicted from theory, we can remove the contribution of ellipsoidal distortion to the overall first harmonic signal by subtracting the expected ellipsoidal modulation. The resultant curve is plotted in red in Figure~\ref{fig:irrad}. Notably, the phasing of this ellipsoidal distortion-corrected signal is almost identical to the measured first harmonic term from $Spitzer$. Since the expected ellipsoidal distortion amplitude at thermal wavelengths is negligible compared to the reported signal from \citet{mansfield2019}, any observed modulation at 4.5~$\mu$m is most likely due to thermal emission from the planet.

The similarity between the ellipsoidal distortion-corrected \tess\ and Spitzer phase curves at the first harmonic lends strong support to the hypothesis that the unusual phase shift we observe is caused by the response of the planet's atmosphere to the time-varying irradiation across its orbit. Of course, the assumption that the star behaves as predicted from equilibrium tide theory is subject to many caveats, as explained earlier, and deviations in the ellipsoidal distortion from expectations will, in turn, affect the corresponding contribution from time-varying irradiance. Future multiwavelength phase-curve observations of KELT-9 and detailed atmospheric modeling of KELT-9b's response to time-varying irradiation can help break the degeneracy between the individual phase and amplitude discrepancies stemming from ellipsoidal distortion and irradiation.


\subsection{Placing KELT-9b in Context}
\label{subsec:context}

\begin{figure}
\includegraphics[width=\linewidth]{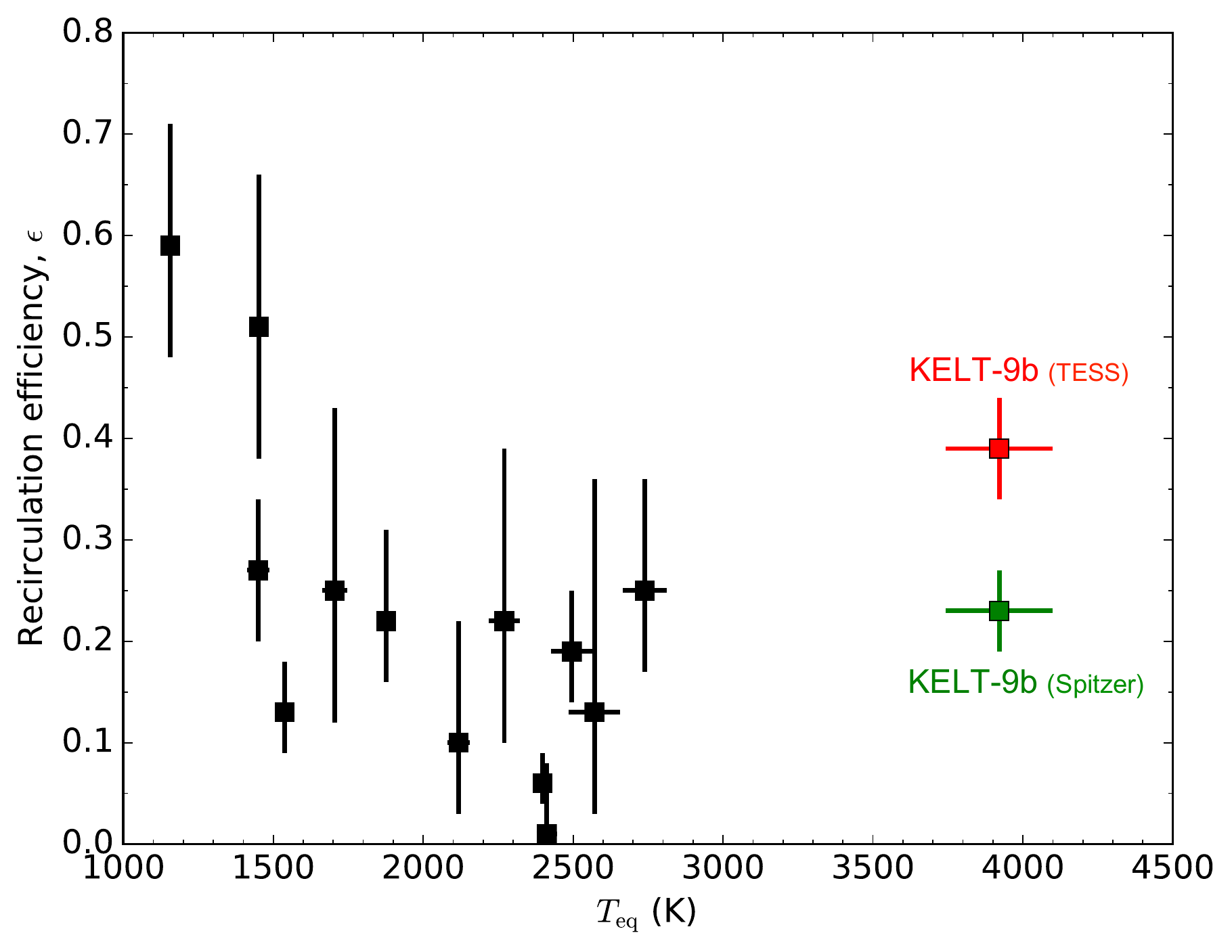}
\singlespace
\caption{ Plot of heat recirculation parameter $\epsilon$ (Equations~\eqref{tday} and \eqref{tnight}) versus dayside equilibrium temperature for 13 transiting objects with published Spitzer phase curves (black points; \citealt{keating2019}). The newly computed $\epsilon$ values derived from the \tess\ and Spitzer phase curves of KELT-9b are plotted in red and green, respectively. While there is significant scatter, the previously reported correlation between increasing dayside equilibrium temperature and decreasing heat transport efficiency is discernible for planets below $T_{\mathrm{eq}}\sim2500~K$. The addition of KELT-9b reveals the turnaround in the trend of day--night heat transport efficiency at around 2500~K, which is consistent with H$_{2}$ dissociation and recombination and its effect on moderating day--night temperature contrasts on the hottest planets.}
\label{fig:eps}
\end{figure}

The growing number of well-measured phase curves has motivated the search for systematic trends that may elucidate the underlying physical processes driving the atmospheric dynamics in exoplanet atmospheres. KELT-9b lies at the high-temperature extreme of the known exoplanets, and we now place the planet into the wider context of hot gas giants.

The most salient trend that has emerged from phase-curve studies is the relationship between day--night temperature contrast and the level of stellar irradiation. As discussed in the Introduction, phase-curve observations have revealed a systematic increase in the day--night temperature contrast with increasing dayside equilibrium temperature up to roughly 2500~K, after which the trend appears to reverse \citep[e.g.,][]{schwartz2017,zhang2018,keating2019}. This behavior is largely consistent with the predictions of general circulation models \citep[e.g.,][]{perna2012,perezbecker2013,komacek2016}. The addition of H$_{2}$ dissociation and recombination to the energy balance of the hottest planets has been invoked to explain the turnaround in the observed trend at $\sim$2500~K \citep{bell2018,komacek2018,tan2019}, though only a handful of planets in this temperature range have previously been observed across the full orbital phase.

Our analysis of the KELT-9b phase curve allows us to test the predictions of the H$_{2}$ recombination hypothesis. \citet{keating2019} calculated heat recirculation parameter values for 13 hot Jupiters and brown dwarfs from Spitzer phase curves, as defined in Section~\ref{subsec:temp}. In Figure~\ref{fig:eps}, we plot the published $\epsilon$ values from their work along with our measurements for KELT-9b. Given the different nightside brightness temperatures and recirculation efficiencies, as well as the possibility that the \tess\ observations probe systematically different pressure levels than Spitzer measurements, we have included the values for both bandpasses. The equilibrium temperature $T_{\mathrm{eq}}=T_{*}\sqrt{R_{*}/2a}$ is used as a proxy for the level of stellar irradiation on the dayside atmosphere. 

\begin{figure}
\includegraphics[width=\linewidth]{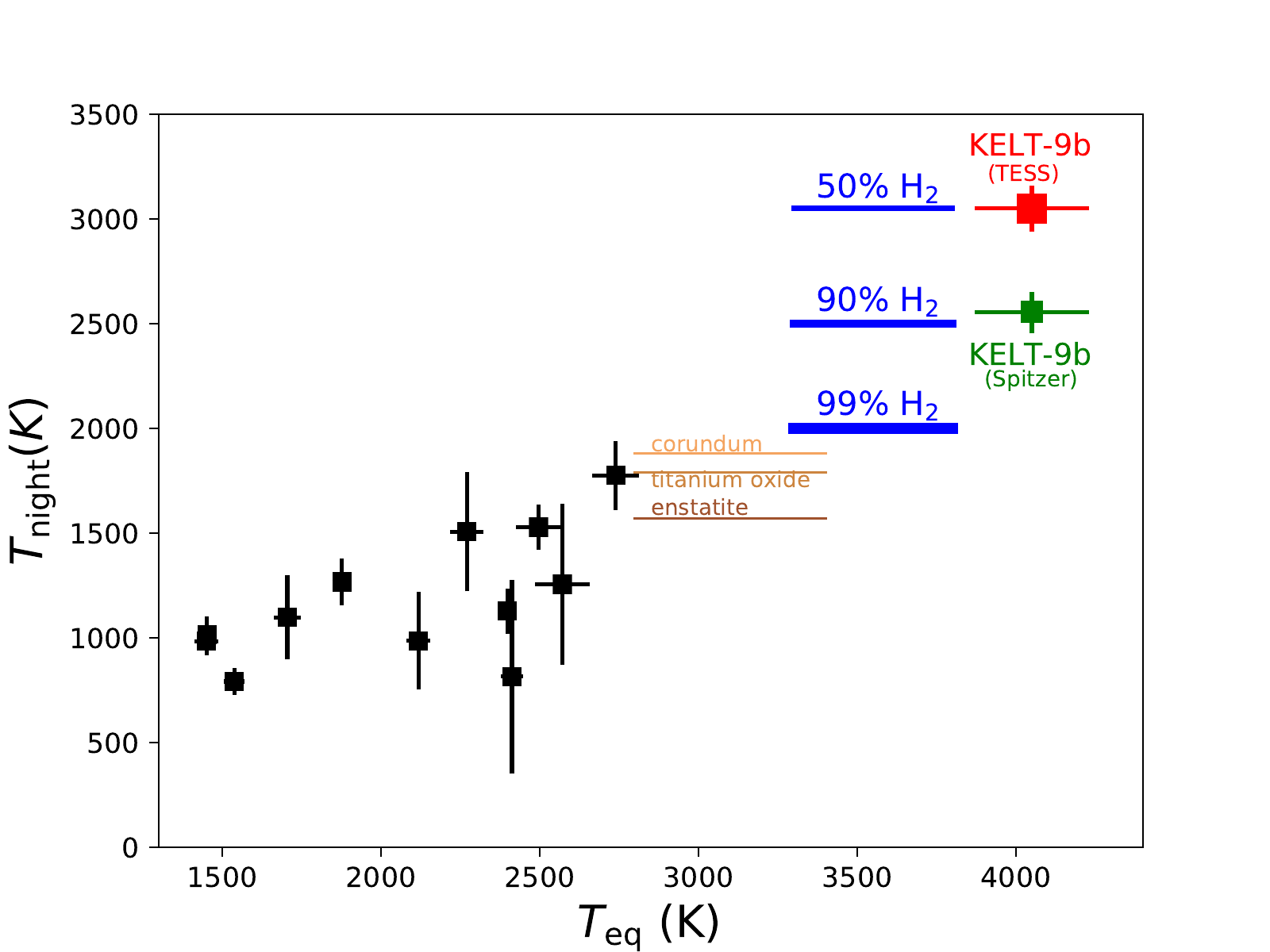}
\singlespace
\caption{Nightside brightness temperatures of hot Jupiters and UHJs with the value obtained in the current study for KELT-9b. Overplotted are the percentages of molecular hydrogen at different temperatures expected from chemical equilibrium in a hydrogen-dominated atmosphere at 0.1~bar. Also overplotted are the condensation temperatures at solar metallicity for enstatite, titanium oxide, and corundum.}
\label{fig:nightside}
\end{figure}


In both bandpasses, the $\epsilon$ value of KELT-9b is elevated relative to the average value for cooler planets with dayside equilibrium temperatures between 2000 and 3000~K. The inclusion of KELT-9b reinforces the previously suggested turnaround in the trend of day--night temperature contrast with irradiation level predicted by the H$_{2}$ recombination hypothesis. Further phase-curve studies of exoplanets are needed, particularly for planets with equilibrium temperatures between 3000 and 3500~K in order to robustly confirm or refute these apparent trends. No such planets have hitherto been discovered.

\cite{keating2019} previously reported a nearly constant nightside temperature of about 1100 K for a sample of 12 hot Jupiters, which they attributed to the presence of opaque nightside clouds. In Figure \ref{fig:nightside}, we have added measured nightside temperatures from \citet{mansfield2019} and our study to the values from \citet{keating2019}. The addition of the KELT-9b data reveals a tentative increasing trend in nightside temperature as a function of insolation across the full dataset. Crucially, the significantly higher nightside temperature of KELT-9b supports the prediction that recombination of H$_{2}$ contributes significantly to nightside heating for the hottest exoplanets \citep{bell2018,komacek2018}.

\subsection{Prospects for Atmospheric Characterization}
\label{subsec:atmo}

In Figure~\ref{fig:nightside}, we have added estimates of the nightside temperature of KELT-9b given various mixing ratios of atomic and molecular hydrogen in the planet's atmosphere, following the methodology of \citet{kitzmann2018}.  Assuming a pure hydrogen gas in chemical equilibrium \citep{gail2014,heng2016} and an approximate photospheric pressure of 0.1~bar consistent with \cite{bell2018} and \cite{kitzmann2018}, the nightside temperature of 3050~K corresponds to an atmosphere with 50\% H$_2$ and 50\% H.  If we consider the full statistical range of nightside temperatures obtained ($3050\pm110$~K), then the percentage of H$_2$ present is 40--60\%.  These estimates place KELT-9b in a very different regime than that considered in \cite{bell2018}, where the nightside is assumed to be H$_2$-dominated.

In Figure~\ref{fig:nightside}, we have also overlaid condensation temperatures at 0.1~bar and solar metallicity for enstatite (MgSiO$_3$), titanium oxide (TiO), and corundum (Al$_2$O$_3$).  Enstatite and corundum are, respectively, the least and most refractory of the mineral clouds that are expected to form in thermochemical equilibrium \citep[e.g.,][]{burrows2016}, while titanium oxide is believed to provide seed particles for high-temperature clouds \citep[e.g.,][]{helling2006}.  The nightside temperatures of KELT-9b in the \tess\ and Spitzer 4.5~$\mu$m bandpasses are $\sim$500--1000~K higher than even the condensation temperature of corundum, implying that the nightside atmosphere is cloud-free at photospheric pressures.

To study the planetary atmosphere in more detail and demonstrate what future observations may reveal, we used the open-source \texttt{HELIOS} radiative transfer code \citep{malik17,malik19} and the \texttt{HELIOS-O} ray-tracing code \citep{gaidos2017} to model the dayside, nightside, and transmission spectra. In this model framework, the emission spectra are computed self-consistently in radiative--convective equilibrium such that the thermal structure is an outcome of the computation, rather than an assumption. We fixed the stellar effective temperature and the planet-to-star radius ratio to the previously published values from \citet{gaudi2017} and \citet{mansfield2019} that we used when fitting for the blackbody brightness temperatures in Section~\ref{subsec:temp}: $T_{\mathrm{eff}}=10{,}170$~K and $R_{p}/R_{*} = 0.08004$. For spectral line lists, we used the ExoMol spectroscopic databases for H$_2$O \citep{barber06,poly18}, CH$_4$ \citep{yt14}, VO \citep{mckemmish2016}, and TiO \citep{mckemmish2019}; for CO, we used \cite{li15}; for CO$_2$, we used the HITEMP database \citep{rothman10}; and spectral lines for Fe, Fe$^+$, Ti, Ti$^+$, Ca, Ca$^+$, Na, K, Y, Y$^+$, AlO, and SH were taken from \citet{Kurucz1995KurCD..23.....K}. In addition, we included the continuum opacities from H$^-$ \citep{john1988} and collision-induced absorption (H$_2$--H$_2$, H$_2$--He, H--He; \citealt{abel11,karman19}), as calculated using the open-source \texttt{HELIOS-K} opacity calculator \citep{gh15}.\footnote{All of the line lists are publicly available at \texttt{http://www.opacity.world}.}  Equilibrium chemistry was computed using the open-source \texttt{FastChem} code, which includes almost 600 chemical species \citep{stock18}.  Solar metallicity was assumed.  For the dayside emission spectra, we computed an interpolated \texttt{PHOENIX} stellar model and used it to irradiate the model atmosphere while varying the heat recirculation parameter $\epsilon$ (as defined in Section~\ref{subsec:temp}). For the nonirradiated nightside, the only free parameter is the interior temperature. In the case of the transmission spectra, we assumed isothermal transit chords.

\begin{figure}
\includegraphics[width=0.9\linewidth]{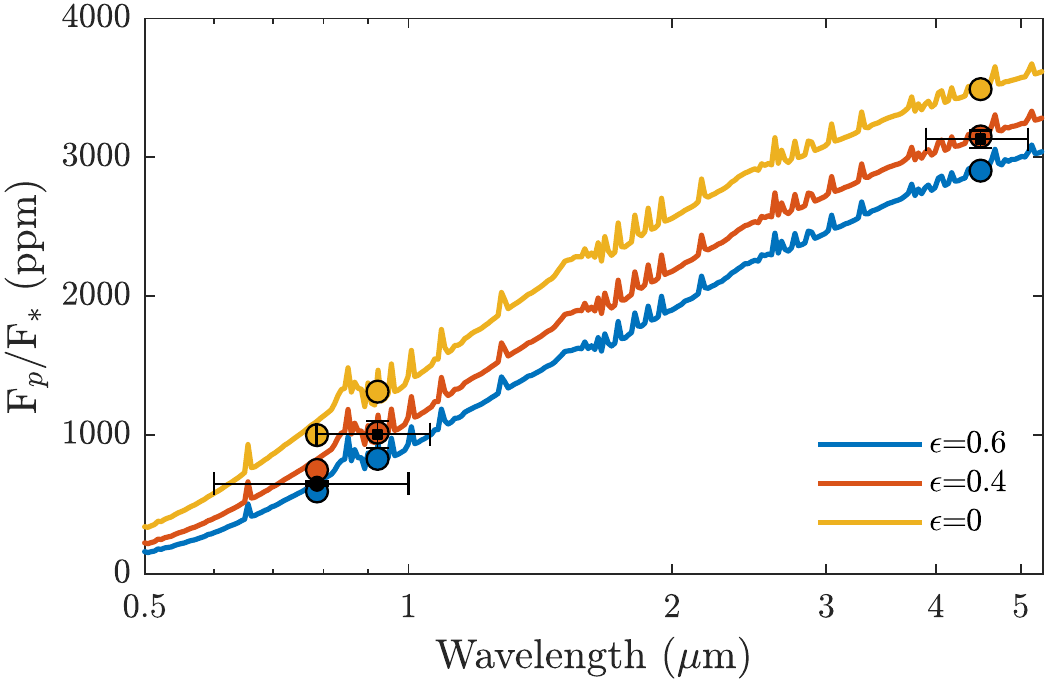}
\includegraphics[width=0.9\linewidth]{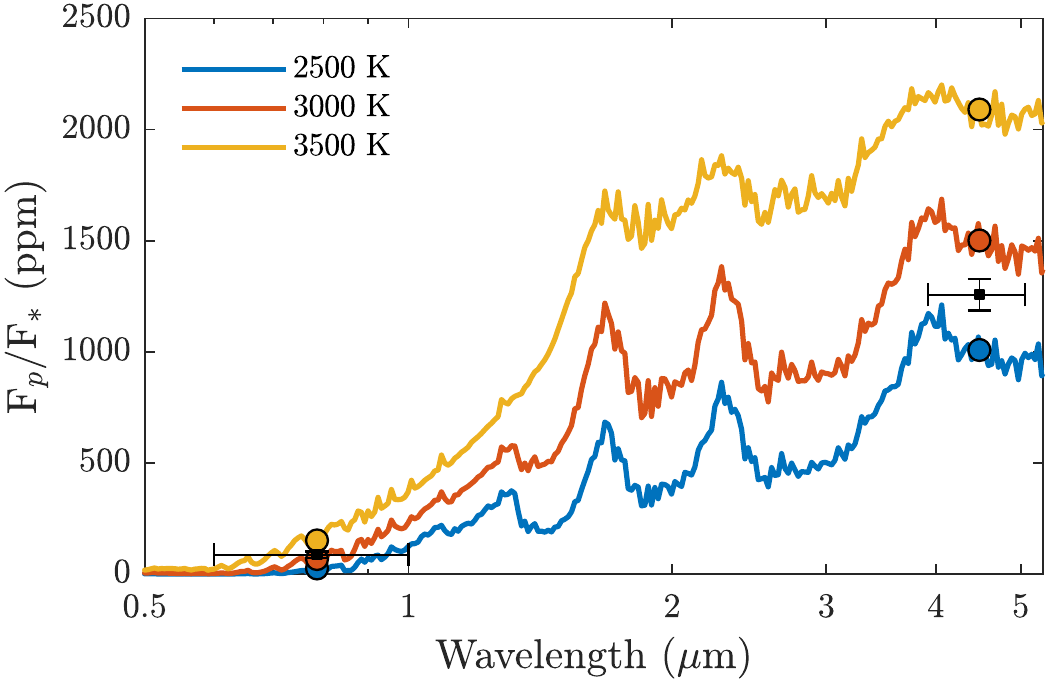}
\includegraphics[width=0.9\linewidth]{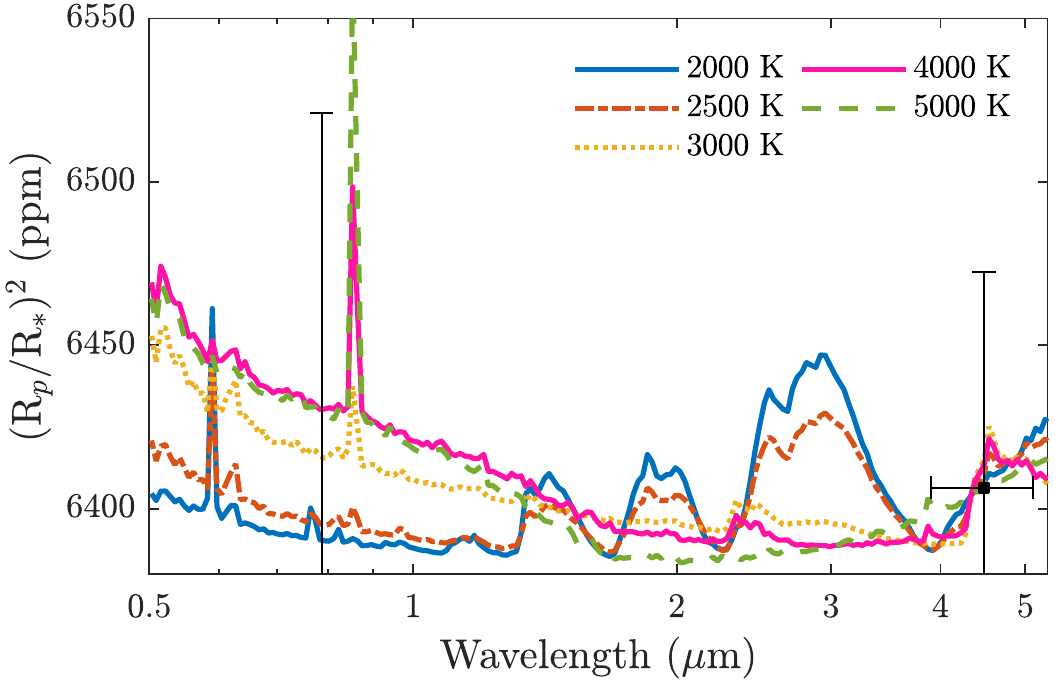}
\singlespace
\caption{Top: predicted 0.5--5~$\mu$m dayside emission spectra from our atmospheric modeling of KELT-9b for various values of the heat recirculation parameter $\epsilon$. Spectral features are from absorption lines in the star's spectrum. All the models are close to a blackbody, and the three published eclipse detections (this paper; \citealt{gaudi2017}; \citealt{mansfield2019}) are best matched by the $\epsilon=0.4$ model (i.e., moderately efficient day--night heat recirculation). Middle: the nightside emission spectra for various assumed values of the interior temperature, showing large absorption features throughout the near-infrared; the measurements from the \tess\ and Spitzer phase curves are overplotted. Bottom: model transmission spectra at various terminator temperatures, with the \tess\ and Spitzer 4.5~$\mu$m transit depth measurements from \citet{ahlers2020} and \citet{mansfield2019} overplotted.}
\label{fig:predictions}
\end{figure}

Figure~\ref{fig:predictions} shows a sampling of models for the dayside emission, nightside emission, and transmission spectra. To facilitate comparison with the measured relative planetary fluxes, the planetary emission spectra are shown as a ratio with the stellar spectrum: $F_{p}/F_{*}$. We have overplotted the measured secondary eclipse depths and nightside fluxes from this paper, as well as the values from \citet{gaudi2017} and \citet{mansfield2019}. For the dayside, we find that the model predicts planetary emission spectra that are close to a blackbody; we note that the spectral features apparent in the modeled $F_{p}/F_{*}$ curves stem from absorption features in the star's spectrum, not the planet's. The three data points are well matched by the model with $\epsilon=0.4$, in excellent agreement with the value of $\epsilon=0.39\pm0.05$ we deduced from thermal balance calculations based on the measured dayside and nightside blackbody brightness temperatures (Section~\ref{subsec:temp}). 

The model nightside emission spectra are shown in the middle panel of Figure~\ref{fig:predictions}. Our measured \tess-band flux is in good agreement with the 3000~K model, while the Spitzer 4.5~$\mu$m flux lies intermediate between the 2500 and 3000~K models. All of the model spectra display large deviations from a blackbody spectrum primarily due to absorption features of H$_{2}$O and CO, as well as the wavelength-dependent H$^{-}$ opacity. In particular, the broad CO absorption in the Spitzer 4.5~$\mu$m bandpass yields a predicted flux that is significantly lower than the corresponding flux when assuming a blackbody. Therefore, this modeling helps explain the discrepant nightside blackbody brightness temperatures measured in the \tess\ and Spitzer 4.5~$\mu$m bandpasses (see Section~\ref{subsec:temp}).

When examining the model--data comparisons, we recall that the forward models shown here assume fixed values for $T_{\mathrm{eff}}$ and $R_{p}/R_{*}$. However, both of these input parameters have relatively large measurement uncertainties, particularly in the case of stellar temperature (450~K). Changes to one or both of these parameters can incur wavelength-dependent variations in $F_{p}/F_{*}$. When experimenting with altering the values of these inputs within their respective $1\sigma$ confidence regions, we found that the relative model brightness at Spitzer wavelengths compared to the \tess-band brightness can vary by $10\%$ or more. In other words, the band-integrated model points shown in the top two panels of Figure~\ref{fig:predictions} have intrinsic error bars that are significant in relation to the distance between adjacent models.

The bottom panel of Figure~\ref{fig:predictions} shows predictions for the transmission spectra of KELT-9b across the wavelength range 0.5--5~$\mu$m. We have chosen to normalize the model spectra to match the Spitzer 4.5~$\mu$m transit depth from \citet{mansfield2019}. We have also included the \tess-band transit depth from \citet{ahlers2020} --- $6260\pm290$~ppm --- which has large error bars but is consistent with the full range of models shown. Meanwhile, we do not include the much larger transit depth from \citet{gaudi2017}, $(R_{p}/R_{*})^2 = 6770\pm70$~ppm; that measurement was derived from transit fits that did not account for the rotational bulge and gravity-darkened limb profile of the rapidly rotating host star (as was done in \citealt{ahlers2020}), despite being in a wavelength range where those effects are significant. 

The most salient trend in the model spectra is the disappearance of the familiar water feature at 1.4~$\mu$m with increasing temperature. As the temperature increases from 2000 to 4000~K, the water abundance decreases by 3 orders of magnitude, while the H$^-$ abundance increases by 2 orders of magnitude (see Figure~2 of \citealt{kitzmann2018}). Given the measured temperatures of KELT-9b, we expect the transmission spectrum at low spectral resolution to be a largely featureless spectral slope across the optical and near-infrared through $\sim$2~$\mu$m, with the exception of the prominent alkali absorption peaks. The gradient of the spectral slope in this wavelength range is determined by the wavelength-dependent behavior of the H$^-$ cross section \citep{john1988}. Meanwhile, at longer wavelengths, our models predict large absorption features, consistent with previous modeling of UHJs \citep[e.g.,][]{lothringer2019}.

The uncertainties in all published broadband transit depths are significantly larger than the range spanned by the modeled transmission spectra. Future spectroscopic observations with the Hubble Space Telescope and/or James Webb Space Telescope, coupled with careful transit shape modeling accounting for the rotationally distorted and gravity-darkened star, may achieve the $\sim$10~ppm precision needed to adequately characterize the transmission spectrum, specifically the spectral gradient in the visible and near-infrared, and allow for a definitive observational test of atmospheric models. Likewise, spectroscopic secondary eclipse observations in the near-infrared and the measurement of absorption features in the dayside emission spectrum will provide strong constraints on the atmospheric chemistry.

\section{Summary}
\label{sec:con}

We have presented a phase-curve analysis of KELT-9 using photometry obtained by the \tess\ mission. The main results of this work are summarized below.
\begin{enumerate}
    \item The secondary eclipse depth of $650^{+14}_{-15}$~ppm implies a dayside brightness temperature of $4600\pm100$~K, assuming zero geometric albedo; for albedos between zero and 0.2, the dayside temperature ranges from 4360 to 4600~K. A simultaneous fit of all secondary eclipse depths in the literature shows that the broadband dayside emission spectrum is consistent with a blackbody at  $4540\pm90$~K and no reflected light. We measure a very hot nightside temperature of $3040\pm100$~K in the \tess\ bandpass, comparable to the atmosphere of a mid-M dwarf.
    \item The planet's atmospheric brightness modulation has a peak-to-peak amplitude of $566\pm16$~ppm and a slight eastward hot-spot offset of $\delta=5\overset{\circ}{.}2\pm0\overset{\circ}{.}9$.
    \item We detect a stellar pulsation signal with a period of $7.58695\pm0.00091$ hr and a peak-to-peak amplitude of $228.1\pm8.5$~ppm.
    \item The photometric variation of the host star is characterized by a strong signal at the first harmonic of the orbital frequency with a semiamplitude of $39.6\pm4.5$~ppm. While the amplitude of this modulation is consistent with the predictions for ellipsoidal distortion, the relative phase of the measured signal lags behind the expected timing by roughly $0.16$ in orbital phase. 
    \item We propose that the discrepant phasing of this signal is likely driven primarily by the additional modulation in thermal emission from the planet's atmosphere in response to the variable irradiation it receives from the gravity-darkened, oblate profile of the rapidly rotating host star over the course of its nearly polar orbit. The orbital geometry and rapid stellar spin, as well as the early-type host star, may also lead to significant deviations in the star's ellipsoidal distortion signal from the predictions of theoretical models.
    \item The relatively small day--night temperature contrast indicates efficient heat transport, confirming the previously suggested turnaround in the trend of heat recirculation efficiency with increasing dayside irradiation level. The muted temperature contrast on KELT-9b is consistent with the effect of H$_{2}$ dissociation and recombination \citep{bell2018,komacek2018}.
    \item The hot nightside temperatures of KELT-9b deviate strongly from the nearly flat trend of nightside temperatures reported by \citet{keating2019}. The elevated nightside temperature precludes the formation of condensate clouds on the planet's nightside and is consistent with an atmospheric composition of roughly equal parts molecular and atomic hydrogen.
    \item The dayside emission spectrum of the planet is predicted to resemble a featureless blackbody. In contrast, the modeled nightside emission spectrum deviates significantly from a blackbody and shows broad molecular absorption features throughout the near-infrared. The 0.5--2.0~$\mu$m transmission spectrum of KELT-9b at low spectral resolution is expected to be largely featureless and resemble a negative spectral slope due to H$^-$ opacity. 
    
\end{enumerate}

\acknowledgments

Funding for the \tess\ mission is provided by NASA's Science Mission directorate. This paper includes data collected by the \tess\ mission that are publicly available from the Mikulski Archive for Space Telescopes (MAST). Resources supporting this work were provided by the NASA High-End Computing (HEC) Program through the NASA Advanced Supercomputing (NAS) Division at Ames Research Center for the production of the SPOC data products. We thank two anonymous referees for their helpful comments that greatly improved the manuscript.
I.W. is supported by a Heising-Simons 51 Pegasi b postdoctoral fellowship. T.D. acknowledges support from MIT’s Kavli Institute as a Kavli postdoctoral fellow. M.M. acknowledges funding from a NASA FINESST grant. Work by J.N.W. was partly supported by the Heising-Simons Foundation.


\end{document}